\documentclass{article}
\pdfoutput=1 

\usepackage{arxiv}

\usepackage[utf8]{inputenc} % allow utf-8 input
\usepackage[T1]{fontenc}    % use 8-bit T1 fonts
\usepackage{hyperref}       % hyperlinks
\usepackage{url}            % simple URL typesetting
\usepackage{booktabs}       % professional-quality tables
\usepackage{amsfonts}       % blackboard math symbols
\usepackage{nicefrac}       % compact symbols for 1/2, etc.
\usepackage{microtype}
\usepackage{booktabs}
\usepackage{graphicx}
\usepackage[ruled,vlined]{algorithm2e}
\usepackage{amsthm}
\usepackage{amsmath}
\usepackage{subcaption}

\newcommand{\norm}[2]{\Vert #1 \Vert_{#2}}

\newcommand{\ip}[2]{\langle #1,#2 \rangle}

\newtheorem{theorem}{Theorem}
\newtheorem{lemma}{Lemma}

\begin{document}
\twocolumn[
\title{Density Sketches for Sampling and Estimation}
\author{Aditya Desai\\
  Department of Computer Science\\
  Rice University\\
  Houston, Texas \\
  \texttt{apd10@rice.edu} \\
   \And
Benjamin Coleman\\
  Department of Electrical and Computer Engineering\\
  Rice University\\
  Houston, Texas \\
  \texttt{brc7@rice.edu} \\
  \And
Anshumali Shrivastava\\
  Department of Computer Science\\
  Rice University\\
  Houston, Texas \\
  \texttt{anshumali@rice.edu} \\
  %% \AND
  %% Coauthor \\
  %% Affiliation \\
  %% Address \\
  %% \texttt{email} \\
  %% \And
  %% Coauthor \\
  %% Affiliation \\
  %% Address \\
  %% \texttt{email} \\
  %% \And
  %% Coauthor \\
  %% Affiliation \\
  %% Address \\
  %% \texttt{email} \\
}
\maketitle
]
\begin{abstract}
We introduce \textit{Density sketches (DS)}: a succinct online summary of the data distribution. DS can accurately estimate point wise probability density. Interestingly, DS also provides a capability to sample unseen novel data from the underlying data distribution. Thus, analogous to popular generative models, DS allows us to succinctly replace the real-data in almost all machine learning pipelines with synthetic examples drawn from the same distribution as the original data. However, unlike generative models, which do not have any statistical guarantees, DS leads to theoretically sound asymptotically converging consistent estimators of the underlying density function. Density sketches also have many appealing properties making them ideal for large-scale distributed applications. DS construction is an online algorithm. The sketches are additive, i.e., the sum of two sketches is the sketch of the combined data. These properties allow data to be collected from distributed sources, compressed into a density sketch, efficiently transmitted in the sketch form to a central server, merged, and re-sampled into a synthetic database for modeling applications. Thus, density sketches can potentially revolutionize how we store, communicate, and distribute data. 

\end{abstract}

%We introduce Density sketches: a succinct summary of the data distribution constructed in a streaming fashion. Density sketches provide accurate pointwise estimates of the probability density and can be used to sample unseen data from the underlying distribution. Modern statistical estimation and modeling applications often require significant amounts of data. However, most techniques allow us to augment or replace the data with synthetic examples, provided they are drawn from the same distribution as the original data. Our lightweight density sketches, with their ability to generate samples, are a perfect alternative to storing terabytes of data for modeling purposes. Density sketches also have many practical advantages, including a streaming algorithm for sketch construction and the ability to merge two sketches of different datasets into a single sketch that summarizes the combined dataset. These properties allow data to be collected from distributed sources, compressed into a density sketch, efficiently transmitted in sketch form to a central server, merged, and resampled into a synthetic database for modeling applications. 

\section{Introduction}
\vspace{-0.2cm}
%\red{need better first paragraph}

The capability to sample from a distribution is a prerequisite for the classical task of statistical estimation, with innumerable applications. The popular Monte Carlo Estimation \cite{montecarlo} algorithm uses sampling as the primary tool to approximate statistical quantities of interest. Data-driven machine learning is yet another example, which is essentially a statistical estimation problem to estimate a model from data. 

Often, we use a set of samples (popularly known as the data) to specify the distribution. Sampling from a distribution described in such a way requires estimating the underlying distribution. Popular methods to infer the distribution and sample from it belong to the following three categories: 1. Parametric density estimation \cite{statlearnbook} 2. Non-parametric estimation - Histograms and Kernel Density Estimators (KDE) \cite{scottbook} 3. Learning-based approaches such as Variational Auto Encoders (VAE), Generative Adversarial Networks (GANs), and related methods \cite{gan,dlbook}. Generally, parametric estimation is not suitable to model most real data as it can lead to large unavoidable bias from the choice of the model~\cite{scottbook}. Learning the distribution, e.g., via neural networks is one solution to this problem. Although learning-based methods have recently found remarkable success, they do not have any theoretical guarantees for the distribution of generated samples. 
%In fact, due to optimization objectives the generated sample is severely biased towards the majority population~\cite{}.
Histograms and KDEs, on the other hand, are theoretically well understood. These statistical estimators of density are known to uniformly converge to the underlying true distribution \textit{almost surely}. In this paper, we focus on such estimators, which have theoretical guarantees. 

\textbf{Our contribution:} Histograms and KDE are the most popular non-parametric density estimation methods. However, they are known to scale poorly with the dimension and size of data. In this work, we propose \textit{Density Sketches}(DS) - a succinct sketch constructed from the data. These sketches are tiny in size and can be used to 1) query the density at a point and 2) sample points from the underlying distribution. We show that Density Sketches are backed by theoretical guarantees and asymptotically approximate true histogram densities. However, unlike histograms, which store counts for exponential (in dimension) number of bins, DS only uses memory logarithmic in size of histogram. Any methodologies for exact computation of the KDE or KDE-based sampling require storing the entire dataset. In contrast, \textit{Density Sketches} do not store the actual data in any form. Furthermore, density sketches have friendly properties that open up a variety of exciting and vital applications. We state some applications below.

\textbf{\textit{Density Sketch} as compressed surrogate for Data}
We propose \textit{Density Sketches}(DS) constructed from data as a compressed alternative to keeping the data itself. Specifically, for statistical estimation tasks, a sample from the underlying distribution is sufficient for the vast majority of applications. In this regard, the original data is not sacrosanct - any sample from the underlying distribution may be used. DS can be used to efficiently sample the required amount of synthetic data from the underlying distribution. With increasing amounts of data, storage and transfer are important practical concerns of growing importance. DS provides an advantageous trade-off for applications.  Our experiments show that with more data, the accuracy of the density sketch improves, but the size is practically unaffected. Our mathematical results show that size is dependent on the variety of the data rather than the volume. These properties make density sketches an appealing alternative to data. 

\textbf{Data Privacy and Sketching}
Fine-grained data collected from many individuals is necessary for many downstream estimation and modeling tasks. However, this could compromise the privacy of an individual. Maintaining the privacy of individuals in released data is an increasingly important topic of study. Data Anonymization is an essential operation towards making data release private, although, with weak privacy guarantees. As \textit{Density Sketches} do not store the exact data, it is a possible way to release anonymized data. Differential privacy \cite{algobook} provides better privacy guarantees for the released data. With some modifications on the lines of \cite{cormodeprivacy}, the \textit{Density Sketch}, which has similar properties to count based sketches, can be made differentially private.

\textbf{Data Collection on Edge and Mobile Devices}
A significant portion of the data today is generated on Edge and Mobile devices. Frequent transfer of data between these devices and a central server is unavoidable because of their limited storage capacity. \textit{Density Sketches}, because of their small size, offers a natural alternative to storing and transferring data on such devices. Density Sketch, being a \textit{sketching} data structure, possesses the mergeable property. It implies that the density sketch of complete data can be achieved by composing density sketches of parts of data. This makes it possible to process data on edge devices to make sketches and combine them at a central location to obtain final \textit{Density Sketch}.

\vspace{-0.2cm}
\section{Background}

%This section cover the topics on which our approach builds. 

\vspace{-0.2cm}
\subsection{Count Sketches}
\vspace{-0.2cm}

Count sketch \cite{cms, cs}, along with its variants, is one of the most popular probabilistic data structure used for the heavy hitter problem. Informally, the heavy hitter problem is to identify top frequent items in a stream of data. This setting can be generalized where each item in data-stream is a key and value pair. Then we can state the problem formally as, given a stream of data of type $(a_t, c_t)$ where $a_t$ is a key belonging to an extensive set, say $\mathcal{U}$, and $c_t$ is the value associated with $a_t$ at time step t. The goal is to output top keys with the most value $c(a)$, where $c(a)$ is the sum of all values associated with key $a$ in the stream.

\begin{figure}[h]
    \vspace{-0.1cm}
    \centering
    \includegraphics[trim= 300 50 300 0 clip,scale=0.3]{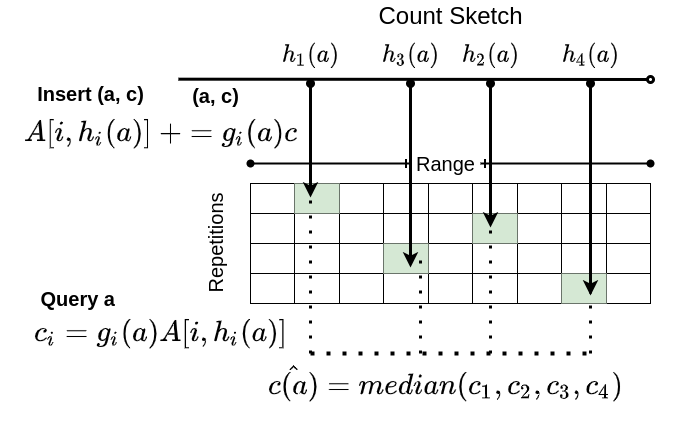}
    \caption{Countsketch, sketching and query}
    \label{fig:cs}
    \vspace{-0.4cm}
\end{figure}

$\mathcal{U}$ can be very large for certain scenarios, for example, to keep track of top queried web pages, the key is web page address which can go up to 100 characters (800bits). In this case $|\mathcal{U}| = 2^{800}$. This prohibits use of array of size $|\mathcal{U}|$. Dictionary to store all keys is prohibitive for its update time and cost of storing all keys. Count sketch offers a probabilistic solution in memory logarithmic in $|\mathcal{U}|$. There is a standard memory accuracy trade-off for count sketches. Let m be the number of distinct keys and $\mathcal{C}$ be the vector of counts indexed by each key. For count median sketch \cite{cs}, the $(\epsilon, \delta)$ guarantee 
\begin{equation*}
    P( |\hat{c(a)} - c(a)| > \epsilon \vert\vert \mathcal{C} \vert \vert_2 )  \leq \delta
\end{equation*}

is achieved using $\mathcal{O}(\frac{1}{\epsilon^2} \frac{1}{\delta} (\log m + \log |\mathcal{U}| ))$ space. \cite{csbook}. As can be seen from the above equation, the approximation accuracy for a particular key depends on how it compares to the $\vert\vert \mathcal{C} \vert \vert_2$. Specifically, count sketch can give very good approximation for keys with highest values in a setting where most of other keys have very low values.

The procedure of sketching and querying in illustrated in figure \ref{fig:cs}. Count sketch, parameterised by K and R, uses K independent pairs of  hash functions ($h_i : \mathcal{U} \rightarrow [0,R]$, $g_i: \mathcal{U} \rightarrow \{-1,1\}$) , $i \in [1,K]$ uniformly drawn from a family of universal hash functions and a 2D array of size $K \times R$, say A. While processing each element (insert operation), say (a, c), for all $i\in[1,K]$, we update $A[i,h_i(a)]$ by adding $g_i(a)c$ to it. To query value c(a), we get K unbiased estimates from each array by querying $A[i, h_i(a)]$ for each i. We can then combine them to get the final estimate of $c(a)$. The median, mean, or median of means are some of the preferred ways of combining them to improve the variance and concentration of estimator around the true value. 
\vspace{-0.2cm}
\subsection{Histograms and Kernel Density Estimation}
\vspace{-0.2cm}
Histograms and KDE \cite{scottbook, scott2004} are popular methods to estimate the density of a distribution given some finite i.i.d sample of size, say n, drawn from the true density, say f(x). 

\textbf{Histogram} divides the support $(S \subset R^d)$ of the data into multiple partitions. It then uses the counts in every partition to predict the density, $\hat{f_H}(x)$, at a point x. Formally the density predicted at the point $x \in S$ is given by 
\begin{align*}
    \hat{f_H}(x) = \frac{c(bin(x))}{n.volume(bin(x))}
\end{align*}
bin(x) identifies the partition in which x lies, c(.) counts the number of samples in that partition, and volume(.) measures the partition volume. Regular Histogram uses hyper cube partitions aligned with data axes with a width, say $h$. $h$ is called the smoothing parameter. As h increases, the estimate's bias increases, and its variance decreases. Histograms suffer from bin-edge problems where a slight change in data across the bin's edge can change predictions. One solution to bin-edge problem is Average Shifted Histogram (ASH) \cite{scottbook}. It uses same bin partitions but with shifted origins. Consider a one dimensional histogram with width h. ASH with m histograms has origins at $0, \frac{h}{m} ,\frac{2h}{m}, ... \frac{(m-1)h}{m}$. The density estimate is then the average of the density estimates obtained from the different histograms. Asymptotically as m goes to $\infty$,  ASH converges to a KDE with the triangle kernel.
%It is easy to see that each micro bin ( of width $\frac{h}{m}$ ) gets a weight depending upon its distance from the point of query. When m tends to $\infty$, the effect of each data point on density becomes apparent. Asymptotically as m goes to $\infty$,  ASH converges to a kernel density estimate with the triangle kernel.

\textbf{Kernel Density} is another smoother estimate of f(x) which resolves the bin-edge problem of histograms. For a given kernel function $k(x,y) :  R^d \times R^d \rightarrow R$ and data, say D, the KDE at point, say x, is defined as 
\begin{equation*}
    \hat{f_K}(x) = KDE(x) = \frac{1}{n}\Sigma_{i\in[1,n], x_i \in D} k(x, x_i)
\end{equation*}
Kernel functions, generally, are positive, symmetric, and integrate to 1. Gaussian, Epanechnikov, Uniform, \cite{statlearnbook} etc. are some of the most widely used kernels. A smoothing parameter h also parameterizes kernel function. It determines its variance for Gaussian kernel, while for uniform and Epanechnikov kernels, it determines the size of the window around x where the function is non-zero. Again, as h increases, the bias increases, and variance decreases.

Histogram and KDEs cannot provide unbiased estimates of the true density \cite{rosenblatt}. Hence, Mean Square Error (MSE) is used to analyze them. Both of these estimators uniformly converge to underlying true distribution asymptotically. However, both suffer from the curse of dimensionality. To get a decent estimate of density in high dimensions, the number of samples needed is exponential in dimensions. Generally, for the Density Estimation task, dimensions of 4-50 are considered large enough. \cite{scott2019}
\vspace{-0.2cm}
\subsection{Randomly Partitioned Histograms and Kernel Density Estimates}
\vspace{-0.2cm}
The density estimate using histogram of a randomly drawn partition is 
\begin{align*}
    \hat{f(x)} \propto \frac{1}{n} \Sigma_{i \in [1,n], x_i \in D} \mathcal{I}(x_i \in bin(x))
\end{align*}
This estimate of the density has a expected value (over random partitions) which looks very similar to a KDE. The subscript p in Expectation is to make it explicit that expectation is over the random partitions.
\begin{align*}
    E_p(\hat{f(x)}) \propto \frac{1}{n} \Sigma_{i \in [1,n], x_i \in D} P(x_i \in bin(x))
\end{align*}
The kernel of this KDE is the probability of collision between the query point x and the data point $x_i$. For example, randomly shifted regular histograms can approximate triangle kernel or asymptotic ASH. Of course, to get a reasonable estimate, we would need to make multiple random partitions and combine estimates obtained from them. Depending on the different partitioning schemes, we can obtain estimators for different kernels. In a recent paper, \cite{race}, authors observe this connection. They show using different lsh functions, it is possible to approximate corresponding interesting kernels. It is useful to note here that if we use l1-lsh or l2-lsh functions \cite{lsh}, the partition is a grid of parallelepipeds of random shape.
 
 \vspace{-0.2cm}
 \subsection{Uniform Sampling from Convex polytopes}
 \vspace{-0.2cm}
 
 Uniform Sampling from Convex spaces is a well-studied problem \cite{hitandrun, vaidya}. For general convex polytopes, this is achieved by finding a point inside the polytope using convex feasibility algorithms and then running an MCMC walk inside the polytope to generate a point with uniform probability. In the case of regular convex polytopes like hypercubes and parallelopiped, uniform sampling is much simpler. Sampling a data point at random in a d-dimensional hypercube of width 1 is equivalent to sampling d real values uniformly in the interval $[0,1]$. For sampling within a $d$-dimensional parallelopiped, we first locate $(d-1)$-dimensional hyperplane parallel to each face at a distance drawn uniformly from $[0,h]$ where h is the width of parallelopiped in that direction. The sampled point is then the intersection of these $(d-1)$ dimensional hyperplanes. 
 
%\subsection{K-Means, Coresets and Random Sample}
%K-Means, coresets and random sample are some of the techniques which can be used to store a good representation of the given data. 
%\paragraph{}
%K-means divides the data into K-clusters and stores the centeroid of each of these clusters. 

\vspace{-0.2cm}
\section{Density Sketches}
%\vspace{-0.2cm}

\begin{table*}[]
\resizebox{\textwidth}{!}{
\begin{tabular}{|c|c|c|l|}
\hline
Partitioning Scheme &
  parameters &
  $bin(x) \in N^d ; x \in R^d$ &
  Sample s from $bin_{id}$ \\ \hline
Regular Histogram &
  width: h &
  $bin(x)_i = \lfloor \frac{x_i}{h} \rfloor$ &
  \begin{tabular}[c]{@{}l@{}}$r \in R^d, r_i \sim Uniform[0,1])$\\ $s = h (bin_{id} + r)$\end{tabular} \\ \hline
Aligned Histogram &
  widths: h = $(h_1, h_2, ... h_d)$ &
  $bin(x)_i = \lfloor \frac{x_i}{h_i} \rfloor$ &
  \begin{tabular}[c]{@{}l@{}}$r \in R^d, r_i \sim Uniform[0,1])$\\ $s_i = h_i (bin_{{id}_i} + r_i)$ \end{tabular} \\ \hline
\multicolumn{1}{|l|}{\begin{tabular}[c]{@{}l@{}}Random Partitions \\using d l1/l2 lsh functions \end{tabular}} &
  \multicolumn{1}{l|}{\begin{tabular}[c]{@{}l@{}} $W \in R^{d \times d} ,W=(w_1, w_2, ..., w_d)$\\ $b : R^{d \times 1},  b = (b_1, b_2, ..., b_d)$\\ width: h\end{tabular}} &
  $bin(x)_i = \lfloor \frac{\ip{x_i}{w_i} + b_i}{h} \rfloor$ &
  \begin{tabular}[c]{@{}l@{}}$r \in R^d, r_i \sim Uniform[0,1])$\\ $y = h * (bin_{id} + r)$\\ Solve $Ws = y-B$ to get s\end{tabular} \\ \hline

\end{tabular}
}
\caption{bin(x) for different partitioning schemes}
\vspace{-0.2cm}
\label{tab:bin}
\vspace{-0.2cm}
\end{table*}

\vspace{-0.2cm}
\subsection{Notation}
\vspace{-0.2cm}
%\begin{itemize}
Data $D$ consists of $n$ i.i.d samples of dimension $d$ drawn from true distribution $f(x) : R^d \rightarrow R$.\newline
\textit{\textbf{bin(x)}}: ID of the partition in which point x falls. In case of regular histograms, RACE style partitioning, $bin(x) : R^d \rightarrow N^d$ and each bin can be identified with a unique tuple of d integers. For example, in regular histogram with width h, $bin(x)_i = \lfloor x_i /h  \rfloor$. Shapes of partitions in the case of regular histograms or RACE style partitioning are regular, and hence simple algorithms can be used for sampling a point inside it. bin(x) and sampling algorithm for some partitioning schemes is mentioned in Table \ref{tab:bin} \newline
\textit{\textbf{mem:}}count sketch with range R and repetitions K as described in section 2.\newline
\textit{\textbf{heap}}: Augmented min-heap of size H used with $mem$.\newline
Hence for a given paritioning scheme $bin(.)$, Density Sketch is parametersized by $(K,R,H) $ and includes two datastructures $mem(K,R)$, $heap(H)$.
\vspace{-0.2cm}
\subsection{Constructing Density Sketches}
\vspace{-0.2cm}

\textbf{Intuition:}
Histogram has an exponential (in d) number of partitions. Hence, we cannot directly build or store a Histogram in high dimensions. However, most high dimensional real data is present in clusters making the Histogram highly sparse in high dimension. So Histogram is an ideal candidate for the heavy-hitter problem. We use count-sketch to store a compressed version of the Histogram. To sample from the Histogram, we sample a bin with probability proportional to its count and then choose a random point in that bin. For an exponential number of bins, this is computationally prohibitive. If we only consider and store heavy bins, this can become achievable. However, getting heavy bins directly from count sketch requires enumerating all possible bins. Hence, we store most heavy bins in a min-heap, which can be updated while data is inserted into the sketch.

As shown in figure\ref{fig:visualization} and algorithm \ref{alg:alg1}%\footnote{Code that implements all the algorithms and used to generate the results in experiments is attached with supplemental material}
, we process the data in a streaming fashion. For each data point, say x, we first find the partition $bin_{id} = bin(x)$ . We increment the count of this $bin_id$ partition by 1 by inserting ($bin_{id}$, 1) into $mem$. Along with each insertion, we also update the heap. If the heap is not at its capacity, we insert this $bin_{id}$ into the heap along with its updated count. In case the heap is at its capacity,  we check $bin_{id}$'s updated count against the minimum of the heap. If $bin_{id}$'s count is found greater, we pop the minimum element from the heap and insert ($bin_{id}$, count). Heap is ordered by the count of the inserted bin ids.

\begin{figure*}
\centering
  \includegraphics[trim=0 50 160 0 clip,width=\textwidth]{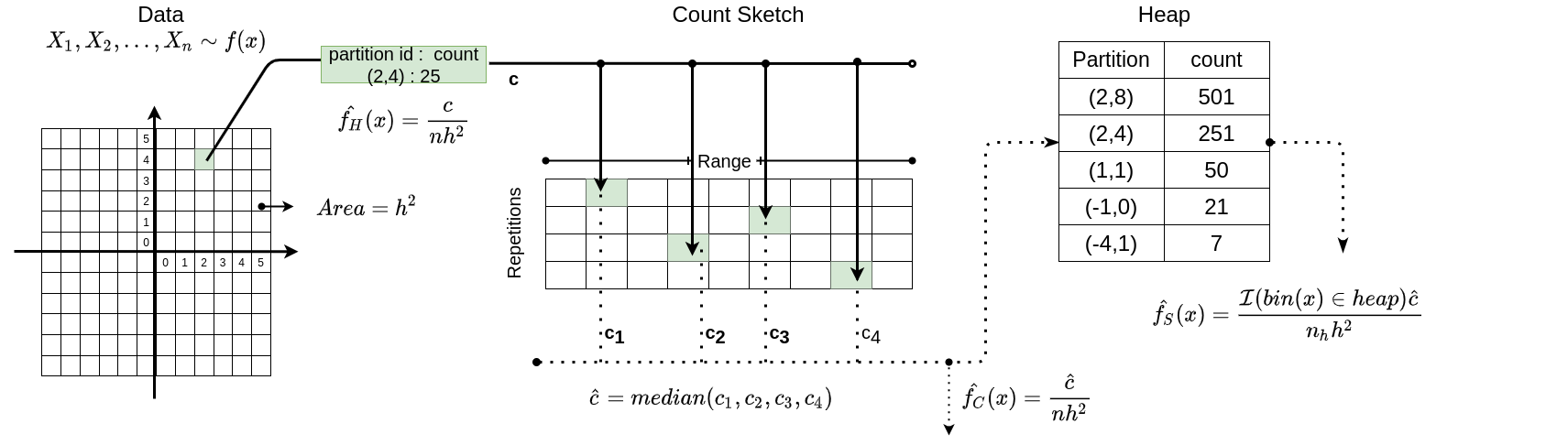}

  \caption{Overview of the sketching algorithm}
  \label{fig:sketching}
  \vspace{-0.6cm}
\end{figure*}

\vspace{-0.2cm}
\subsection{$\hat{f_C}(x)$: Estimate of density at a point}
\vspace{-0.2cm}
We can use these sketches for querying the density estimate at a particular point. The algorithm for querying is presented in Algorithm \ref{alg:alg2} and is explained in the figure \ref{fig:sketching}. The density predicted by the histogram, say $\hat{f_H}(x)$, can be written as,
\begin{align*}
    &\hat{f_H}(x) = \frac{c(bin(x))}{n. volume(bin(x))}\\
    &volume(bin(x)) = h^d \quad (regular\ histogram)
\end{align*}
where c(bin(x)) is the count of data points that lie in bin(x). When using the sketch, instead of using actual c(bin(x)), we would use the estimate of c(bin(x)) from the count sketch. Let this estimate be $\hat{c}(bin(x))$. Then we can write the density predicted using countsketch as $\hat{f_C}(x)$
\begin{align*}
    \hat{f_C}(x) = \frac{\hat{c(bin(x))}}{n. volume(bin(x))}
\end{align*}
 We know from countsketch literature, that $\hat{c(x)}$ is closely distributed around c(x) and so we can expect $\hat{f_C}(x)$ to be close to $\hat{f_H}(x)$ and hence to $f(x)$. Note that though $\hat{f_C}(x)$ is a good estimate of density at a point x, the function $\hat{f_C}(.)$ is not a density function as it does not integrate to 1. 

\begin{algorithm}
\SetAlgoLined
\KwResult{Density Sketch}
 $f(x) : R^d \rightarrow R$ : true distribution\\
 $x_1, x_2, \ldots x_n \sim f(x) $ : sample drawn from $f(x)$ \\
 bin(x) : $R^d \rightarrow N^d$:  bin to which x belongs \\
 mem : CountSketch(R, K) : sketch with R-range, K-repetitions \\
 heap : Heap(H): min-heap to store top H elements \\
 \For{i $\gets$ 1 $\KwTo$ n} {
    $bin_{id} = bin(x_i)$\\
    $mem.insert(bin_{id}, 1)$\\
    $c = mem.query(bin_{id})$\\
    $heap.update(bin_{id}, c)$
 }
\caption{Constructing density sketch of $f(x)$}
 \label{alg:alg1}
\end{algorithm}
\vspace{-0.5cm}
\begin{algorithm}
\SetAlgoLined
\KwResult{$\hat{f_C}(y)$}
 $y \in R^d$\\
 $bin_{id} = bin(y)$\\
 $count = mem.query(bin_{id})$\\
 return $\frac{count}{n Volume(bin_{id})}$
 \caption{query $\hat{f_C}(y)$, $y \in R^d$}
 \label{alg:alg2}
\end{algorithm}
\vspace{-0.5cm}
\subsection{$\hat{f^*_C}(x)$: Estimate of density Function}
\vspace{-0.2cm}

In order to obtain a density function from the sketches, we have to normalize the function $\hat{f_c}(x)$ over the support. We can write $\hat{f^*_C}(x)$ as 
\begin{align*}
    \hat{f^*_C}(x) &\propto \hat{c}(x) \quad \hat{f^*_C}(x) =\frac{\hat{c}(x)}{\int \hat{c}(x) dx}
\end{align*}

It is easy to check the integral can be written as the sum over all the bins in the support. 

\begin{align*}
    \hat{f^*_C}(x) &=\frac{\hat{c}(x)}{\textrm{volume}(bin(x)) \Sigma_{b \in bins} \hat{c}(b)}\\
                  &=\frac{\hat{c}(x)}{\textrm{volume}(bin(x)) \hat{n}}
\end{align*}
As is clear from the equations for $\hat{f^*_C}(x)$ and $\hat{f_C}(x)$, $n = \Sigma_{b \in bins} c(b)$ , is replaced by $\hat{n} = \Sigma_{b \in bins} \hat{c}(b)$ to get a density function. We can check that $\hat{n}$ is an estimate of n using estimate of count for each bin from the density sketch. 
\vspace{-0.2cm}
\subsection{$\hat{f_S(x)}$: Sampling from Density Sketches}\label{sec:sampling}
\vspace{-0.2cm}

The count sketch is a good enough representation for querying the density at a point. However, it is not the best data structure to efficiently generate samples. One naive way of sampling from these sketches is to randomly select a point in support of f(x) and then do a rejection sampling using estimate $\hat{f_C}(x)$. However, given the enormous volume of support in high dimensions, this method is bound to be immensely inefficient. Another way is to choose a partition with probability proportional to the count of elements in that partition and then sample a random point from this chosen partition. It is easy to check that the probability of sampling a point x in this manner, precisely, is $\hat{f_H}(x)$ if we use exact counts and $\hat{f^*_C}(x)$ if we use approximate counts from count sketch. However, given that number of bins is exponential in dimension, sampling a bin proportional to its counts requires prohibitive memory and computation. This is essentially the reason why we needed a count sketch in the first place.  Here, we further approximate the distribution by storing only top H partitions which contain most data points and discarding other partitions.  As mentioned in \ref{alg:alg1}, we can maintain top H partitions efficiently with an augmented heap. We then sample a partition present in this heap with probability proportional to its count and sample a random data point from this partition (Algorithm \ref{alg:alg3}). The probability of sampling a data point whose bin is not present augmented heap is then zero. The distribution of this sampling algorithm is, 
\begin{align*}
    &\hat{f_S}(x) = \mathcal{I}(bin(x) \in heap) \frac{\hat{c}}{\hat{n_h} volume(bin(x))}
\end{align*}

where $\hat{n_h} = \Sigma_{b \in heap} \hat{c}(b)$, is the count-sketch estimate of total number of elements captured in all partitions present in heap. $\mathcal{I}(.)$ is the indicator function with values 0 or 1 evaluating the Boolean statement inside it. Let $ratio_h = \hat{n_h} / \hat{n}$ be the capture ratio of heap. It is easy to see that as capture ratio tends to 1, $\hat{f_S}(x)$ tends to $\hat{f^*_C}(x)$. Note that $\hat{f_S}(x)$ is indeed a density function.

\begin{algorithm}
\SetAlgoLined
\KwResult{y: sample from $f_S(y)$}
 P : multinomial distribution, such that\\
 -- $P(bin_{id}) = \frac{\hat{c(bin_{id})}}{n_h} \textrm{ if } bin_{id} \in heap.$ \\
 -- $P(bin_{id}) = 0 \textrm{ if } bin_{id} \notin heap. $\\
 $bin_{id} \sim P $\\
 $y = UniformRandom(partition(bin_{id}))$

\caption{sample $y \in R^d$ such $y \sim \hat{f_S}$ }
 \label{alg:alg3}
\end{algorithm}
\vspace{-0.2cm}

\vspace{-0.2cm}
\section{Analysis}
\vspace{-0.2cm}
Histogram and Kernel Density Estimators are  well studied non-parametric estimators of density. Both of these estimators are shown to be capable of approximating a large class of  functions \cite{scottbook}.  For example , with the condition of Lipschitz Continuity on f, we can prove that point wise $MSE(\hat{f_H}(x)$ converges to 0 at a rate of $\mathcal{O}(n^{-2/3})$.  Better results can be obtained for functions that have continuous derivatives.  In our analysis we make all such assumptions on the lines of those made in \cite{scottbook}; specifically, existence and boundedness of all function dependent terms that appear in the theorems below. We refer reader to \cite{scottbook} for in depth discussion on assumptions.

 We restrict our analysis to convergence in probability for all the estimators discussed in this paper which is standard \cite{scottbook}. In this section we consider the regular histogram partitioning scheme and show that our density estimates $\hat{f^*_C}(x)$ and sampling distribution $\hat{f_S}(x)$ are approximations of underlying distribution f(x) and converge to it. However, similar analysis holds even for random partitioning schemes / KDE and is skipped here.
 
\textbf{Mean integrated square error(MISE)}: MISE of an estimator of function is a widely used tool to analyse the performance of a density estimator.
\begin{equation*}
    MISE = E \int (\hat{f}(x) - f(x))^2 dx
\end{equation*}
A density estimator, with MISE asymptotically tending to 0, is a consistent estimator of true density and converges to it in probability. We would use this tool to make statements about convergence of our estimators. By Fubini's theorem, MISE is equal to IMSE (Integrated mean square error. Also, for any estimator $\hat{e}$, $MSE(\hat{e}) = Variance(\hat{e}) + Bias(\hat{e})^2$. Hence we can write IMSE  (and hence MISE) as sum of integrated variance (IV) and integrated square bias (ISB).
\begin{align*}
    &IMSE(\hat{f}) = \int E((\hat{f}(x) - f(x))^2) dx = IV + ISB
\end{align*}
The remaining section is organized as follows. We first state the main theorem in our paper which relates the Sampling probability of Density Sketches, $\hat{f_s}(x)$, to the underlying true distribution. We then state various theorems that interrelate the different density function estimates used in this paper. The combination of all these theorems lead to our main theorem. The proofs of all the theorems can be found in Appendix A. We then briefly interpret different theorems in a small subsection.

\begin{theorem}[\textbf{Main Theorem}$\hat{f}_S(x)$ to $f(x)$ ]
The probability density function of sampling, $\hat{f_s}(x)$, using a Density Sketch over regular histogram of width h, with parameters(K,R,H) created with n i.i.d samples from original density function f(x), has an IMSE :
\begin{align*}
IMSE(&\hat{f_S}(x)) \leq 12(1-ratio_h)^2 \\ &+3 (1+2\epsilon) (\frac{1}{nh^d} + \frac{R(f)}{n} + o(\frac{1}{n}) + \frac{\#bins - 1}{KRnh^d}) \\
&+3(1+ 3\epsilon) \frac{h^2d}{4} R(\norm{\nabla f}{2}) \\
 &+3 \epsilon (1 + 2 \mathcal{R}(f) + h\sqrt{d} \int_{x \in S}(f(x) \norm{\nabla f}{2})))
\end{align*}
with probability $(1-\delta)$ , where $\delta = \frac{\#bins }{\epsilon^2 n R}$, $\#bins$ is the number of non-empty bins in histogram, $ratio_h$ is the estimated capture ratio as described in section \ref{sec:sampling}
\end{theorem}

The dependence of IMSE on properties of f(x), such as roughness, is standard \cite{scottbook} and cannot be avoided. 

\textbf{Interpretation} The estimator $f_S(x)$ of $f(x)$ is obtained by a series of approximations from $f(x) \rightarrow f_H(x) \rightarrow f_C(x) \rightarrow f^*_C(x) \rightarrow f_S(x)$. Hence in order to interpret this result, we break down the result above into multiple theorems enabling the reader to easily notice which step of approximations lead to what terms in the theorem above. We urge readers to read the interpretation of each theorem below to obtain a complete picture of the Main theorem.

\begin{theorem} $[\mathbf{f_H(x) \textrm{ to } f(x)}]$
The IMSE for the estimator $\hat{f_H}(x)$ using  regular histogram with width h built over n i.i.d samples drawn from true distribution f(x), is 
\[
    IMSE(\hat{f_H}) \leq \frac{1}{nh^d} + \frac{R(f)}{n} + o(\frac{1}{n}) + \frac{h^2d}{4} R(\norm{\nabla f}{2})
\]
Specifically, its $IV = (\frac{1}{nh^d} + \frac{\mathcal{R}(f)}{n} + o(\frac{1}{n}))$ and  $ISB \leq (\frac{h^2d}{4} \mathcal{R}(\norm{\nabla f}{2}))$ where $\mathcal{R}(\phi)$ is the roughness of the function $\phi$ defined as $\mathcal{R}(\phi) = \int_{x \in S} \phi(x)^2 dx$
\end{theorem}
\textbf{Interpretation }: Theorem 2 applies to all the functions for which we can apply Taylor series expansion up to 2 terms and roughness terms R(f) and $R(\norm{\nabla f}{2})$ exist and are bounded. It is clear that if $h \rightarrow 0$ and $nh^d \rightarrow \infty$, then $IMSE \rightarrow 0$, which implies that $\hat{f_H}(x)$ converges to f(x) asymptotically. As $nh^d \rightarrow \infty$, n should grow at a rate faster than reduction of $h^d$. This is exactly the curse of dimensionality. 

\begin{theorem} $[\mathbf{f_C(x) \textrm{ to } f(x)}]$
The IMSE of estimator $\hat{f_C}(x)$ obtained from the Density Sketch with parameters(R,K,$\_$) using histogram of width h built over n i.i.d samples drawn from true distribution f(x) is 
\[
    IMSE(\hat{f_C}) = IMSE(\hat{f_H}) + \frac{\#bins - 1}{KRnh^d}
\]
where $\#bins$ is the number of non-zero bins in histogram. Specifically, its  $IV(\hat{f_C}) = IV(\hat{f_H}) + \frac{\#bins - 1}{KRnh^d} $ and $ISB(\hat{f_C}) = ISB(\hat{f_H})$
\end{theorem}
\textbf{Interpretation:} Note that though $\hat{f_C}(x)$ is not a density estimator as $\int \hat{f_C}(x) \neq 1$ , it is still useful to look at the IMSE for this function. It is clear from the above theorem, if $\frac{\#bins}{KR}$ is bounded, then as $h \rightarrow 0$ and $nh^d \rightarrow \infty$, $IMSE(\hat{f_C})$ tends to 0 and the $\hat{f_C}(x)$ converges to f(x) simultaneously at all points in probability. As expected, using count sketches adds to the variance of the estimator while keeping the bias unchanged. The term $\frac{\#bins}{KR}$ gives the memory accuracy trade-off here, as it compares the number of keys being inserted into sketch versus the Density Sketch memory (=KR) excluding heap.
\begin{figure*}

\begin{subfigure}{.33\textwidth}
  \centering
  \includegraphics[trim=100 0 100 0,clip,width=.99\linewidth]{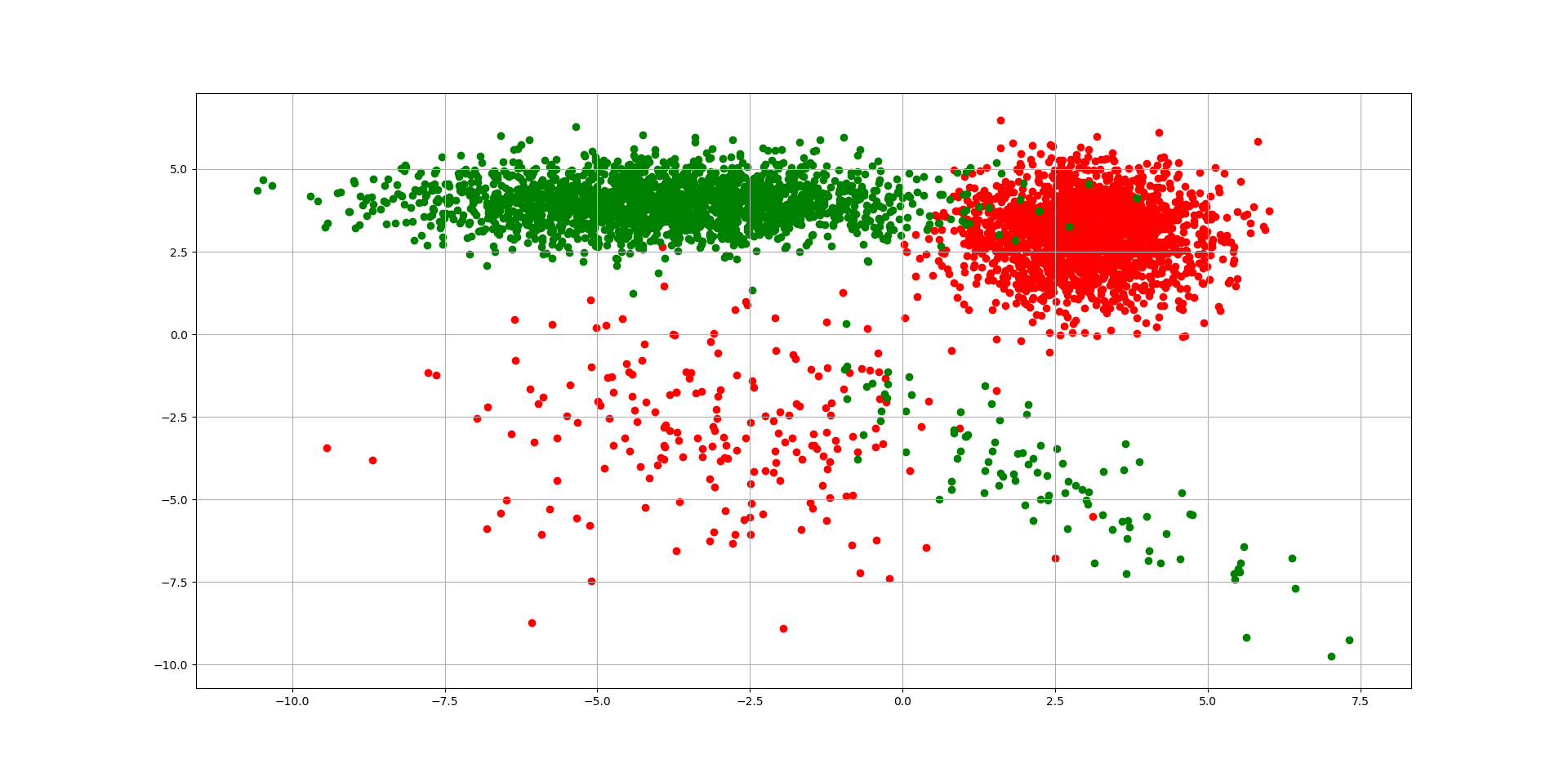}
  \caption{ sample drawn from true distribution }
 
\end{subfigure}
\begin{subfigure}{.33\textwidth}
  \centering
  \includegraphics[trim=100 0 100 0,clip,width=.99\linewidth]{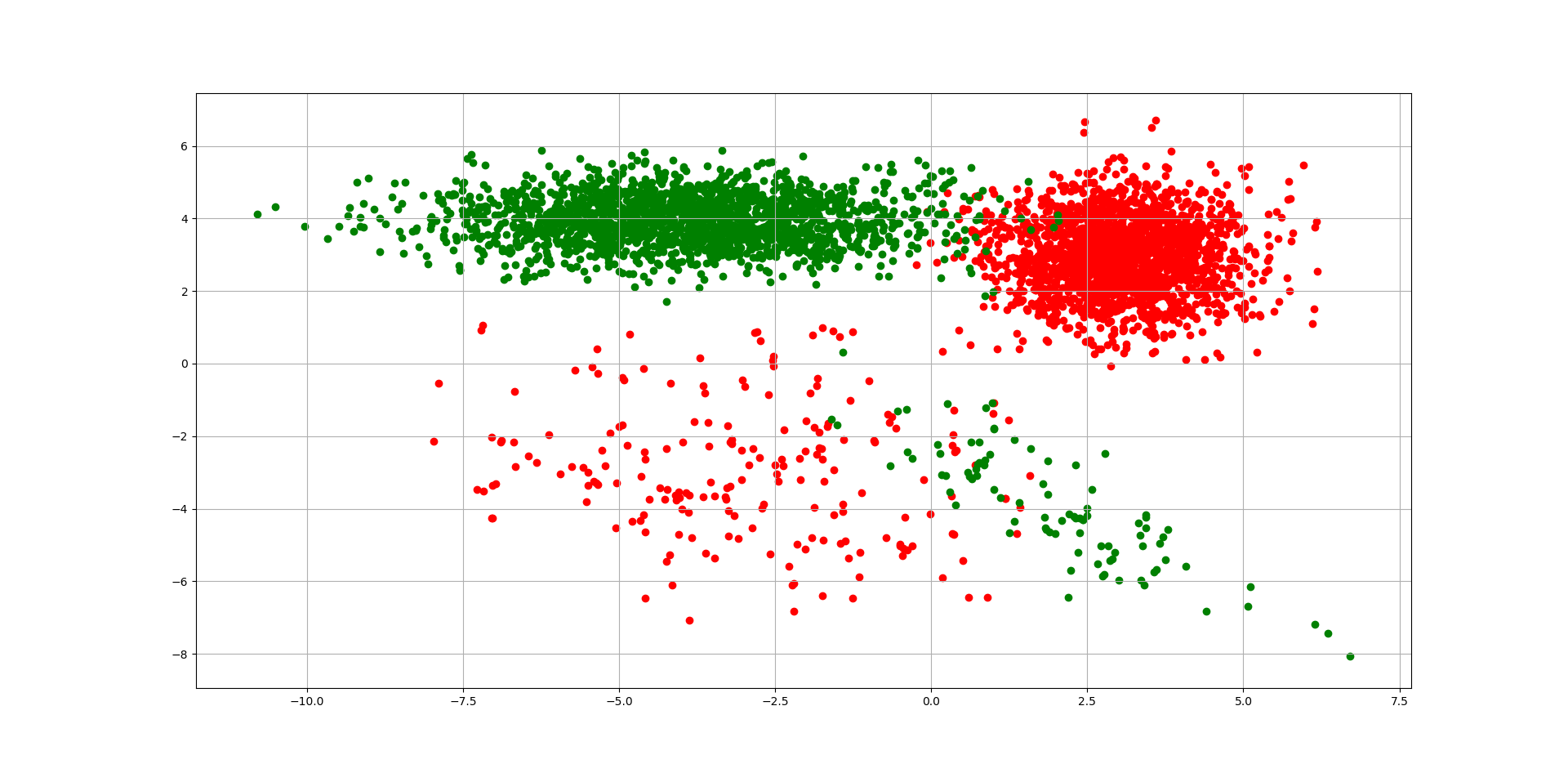}
  \caption{ sample drawn from Density Sketch }

\end{subfigure}
\begin{subfigure}{.33\textwidth}
  \centering
  \includegraphics[trim=100 0 100 0,clip,width=.99\linewidth]{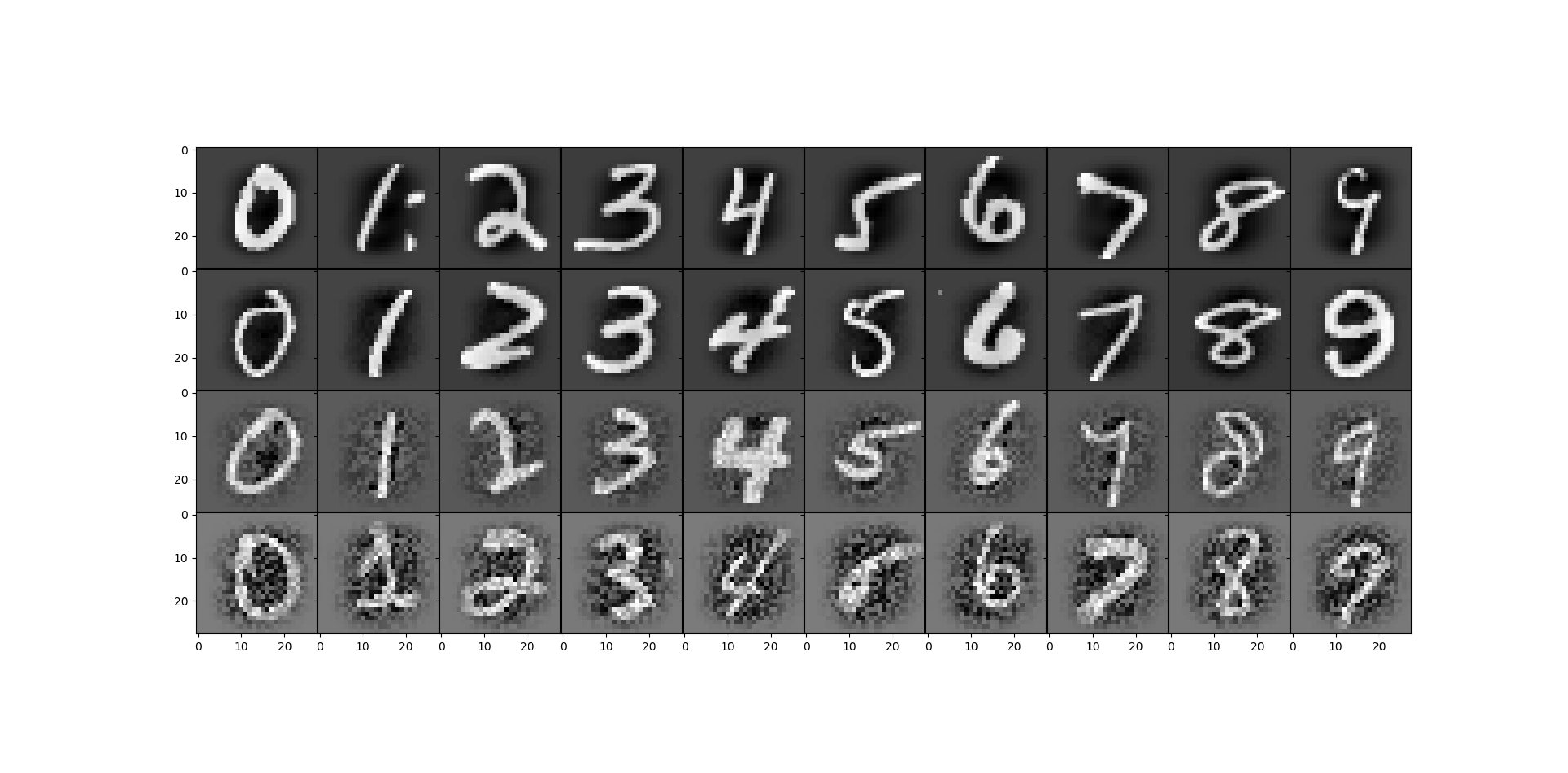}
  \caption{ MNIST Sample from Density Sketch using width: 0.01,0.1,1,10 from top to bottom }
  
\end{subfigure}
\caption{Visualization of samples drawn from Density Sketch}
\label{fig:visualization}
\vspace{-0.6cm}
\end{figure*}

\begin{theorem}$[\mathbf{f^*_C(x) \textrm{ to } f_C(x)}]$
The IMSE of estimator $\hat{f^*_C}(x)$ obtained from the Density Sketch with parameters(R,K,$\_$) using histogram of width h built over n i.i.d samples drawn from true distribution f(x) is 
\[
     |IMSE(\hat{f^*_C}) - IMSE(\hat{f_C})| \leq  \epsilon (N + 2M)
\]
\begin{align*}
\textrm{Specifically,}   &  \quad  \quad |IV(\hat{f^*_C}) - IV(\hat{f_C}) | \leq  2 \epsilon M\\
                        & \quad  \quad |ISB(\hat{f^*_C}) -  ISB(\hat{f_C})| \leq  \epsilon N\\
 \textrm{where,}  & \quad  \quad N = (1 + ISB(\hat{f_C})) \\ 
                  & \quad  \quad M =  IV(\hat{f_C}) + 2 \mathcal{R}(f) + (h^2d/4) \mathcal{R}(\norm{\nabla f}{2}) \\
                  & \quad  \quad + h\sqrt{d} \int_{x \in S}(f(x) \norm{\nabla f}{2}) )
\end{align*}
with probability $(1 - \delta)$ where $\delta = \frac{\#bins }{\epsilon^2 n R}$
\end{theorem}

\textbf{Interpretation:} We provide a probabilistic $(\epsilon, \delta)$ guarantee on how $\hat{f^*_C}(x)$ approaches $\hat{f_C}(x)$ as the parameters of Density Sketch  and number of data points change. The $\epsilon, \delta$ give the accuracy-memory tradeoff that occurs due to restricted memory of the Density Sketch (excluding heap). Using sufficiently large R, we can essentially control the deviance of $\hat{f^*_C}$ from $\hat{f_C}$.

%We now look at the probability of the sample generated by our algorithm $\hat{f_S}(x)$. $\hat{f_S}(x)$ is parameterised by the heap size H. As the heap size H increases, the capture ratio , $ratio_h = \frac{\hat{n_h}}{\hat{n}}$ also increases. Specifically, when $H = \#bins$, i.e. heap size is equal to number of non-empty partitions, $ratio_h$ will be 1. We will state the following theorem connecting $\hat{f_S}(x)$ and $\hat{f^*_C}(x)$ in terms of the capture ratio instead of H. However, it should be clear that $ratio_h$ is data dependent and we control $ratio_h$ via H.
\begin{lemma}$[\mathbf{f_S(x) \textrm{ to } f^*_C(x)}]$
Estimators $\hat{f_S}(x)$ and $\hat{f^*_C}(x)$,  obtained from the Density Sketch with parameters(R,K,H) using histogram of width h built over n i.i.d samples drawn from true distribution have a relation
\[
\int |\hat{f^*_C}(x) - \hat{f_S}(x)| dx = 2(1-ratio_h)
\]
where $ratio_h$ is the capture ratio as defined in \ref{sec:sampling}
\end{lemma} 
Using above lemma, we get a bound on the IMSE as follows
\begin{theorem}$[\mathbf{f_S(x) \textrm{ to } f^*_C(x)}]$
The IMSE of estimator $\hat{f_S}(x)$ obtained from the Density Sketch with parameters(R,K,H) using histogram of width h built over n i.i.d samples drawn from true distribution f(x) is 
\[
IMSE(\hat{f_S}(x)) \leq 12(1-ratio_h)^2 + 3 IMSE(\hat{f^*_C}(x))
\]
where $ratio_h$ is the capture ratio as defined in \ref{sec:sampling}
\end{theorem}
\vspace{-0.2cm}
\begin{table}
\resizebox{\linewidth}{!}{
\begin{tabular}{|l|l|l|l|}
\hline
Dataset                                                     & Dimension & Samples & zipped size \\ \hline
skin/nonskin                                                & 3         & 180K    & 670KB       \\ \hline
susy                                                        & 18        & 4.5M    & 0.8GB       \\ \hline
higgs                                                       & 28        & 10M     & 2.5GB       \\ \hline
webspam(unigram)                                                   & 254       & 300K    & 120MB       \\ \hline
\end{tabular}
}
\caption{Classification Datasets from \cite{liblinear} with dimenstion $< 500$, number of samples per class $> 100,000$}
\label{tab:datasets}
\vspace{-0.1cm}
\end{table}
\textbf{Interpretation} From Lemma 1, we can infer that for sufficiently large H, when $ratio_h = 1$,  $\hat{f^*_C}(x)$ and $\hat{f_S}(x)$ are exactly the same simultaneously at all points in the support. Specifically $ratio_h$ for a particular H, captures the dependence of accuracy of $\hat{f_S}(x)$ upto $\hat{f_C}(x)$ on the data. If the data is clustered and stored in various pockets of the space, heap captures almost all of the data, $ratio_h$ would essentially be close to 1 suggesting that $\hat{f_S}$ will be close to $f^*_C$. Similar behaviour would be seen if the distribution of data into bins would follow a power law; implying that most data is concentrated in few bins which will be captured in the heap. If the data is scattered then $ratio_h$ would be worse and hence the two estimators would diverge.

\begin{figure*}
\begin{subfigure}{.24\textwidth}
  \centering
  \includegraphics[width=.98\linewidth]{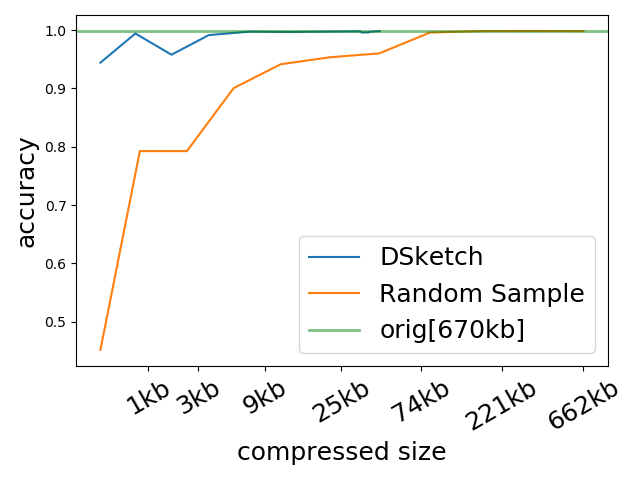}
  %\caption{skin/nonskin}
\end{subfigure}
\begin{subfigure}{.24\textwidth}
  \centering
  \includegraphics[width=.98\linewidth]{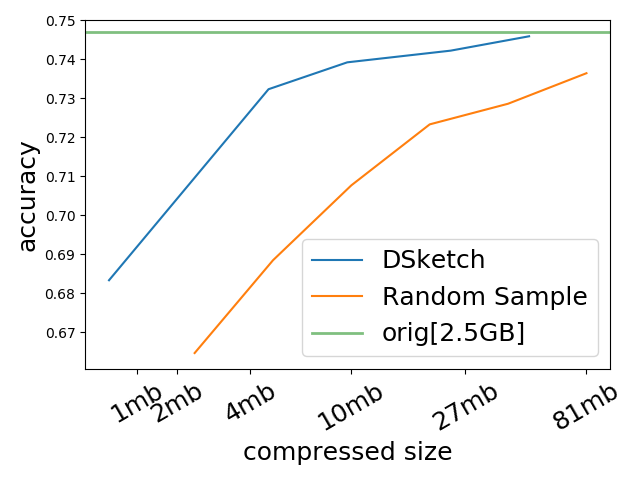}
  %\caption{higgs }
\end{subfigure}
\begin{subfigure}{.24\textwidth}
  \centering
  \includegraphics[width=.98\linewidth]{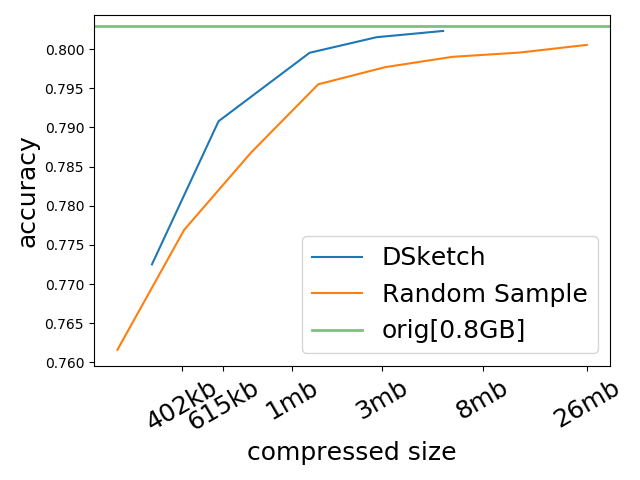}
  %\caption{ susy}
\end{subfigure}
\begin{subfigure}{.24\textwidth}
  \centering
  \includegraphics[width=.98\linewidth]{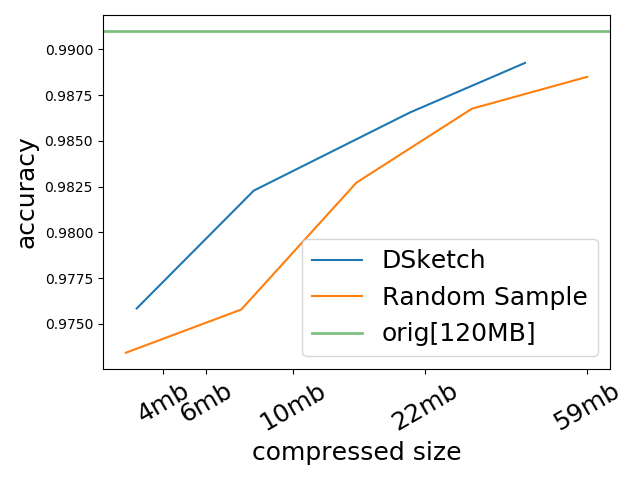}
  %\caption{webspam(unigram)}
\end{subfigure}

\label{fig:classification}
%\end{figure} 
%\begin{figure}
\begin{subfigure}{.24\textwidth}
  \centering
  \includegraphics[width=.98\linewidth]{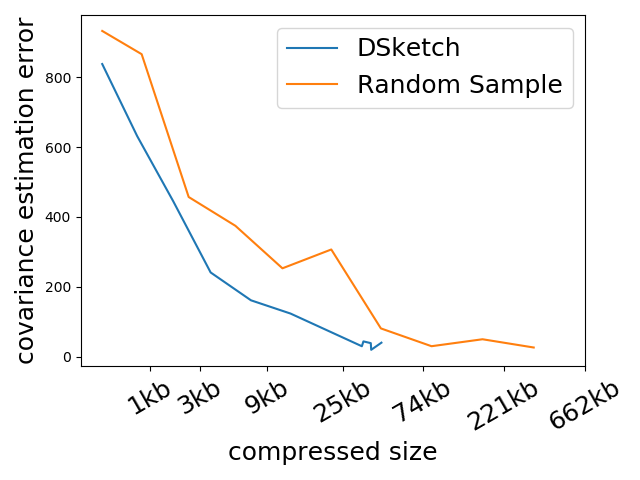}
  \caption{skin/nonskin}
\end{subfigure}
\begin{subfigure}{.24\textwidth}
  \centering
  \includegraphics[width=.98\linewidth]{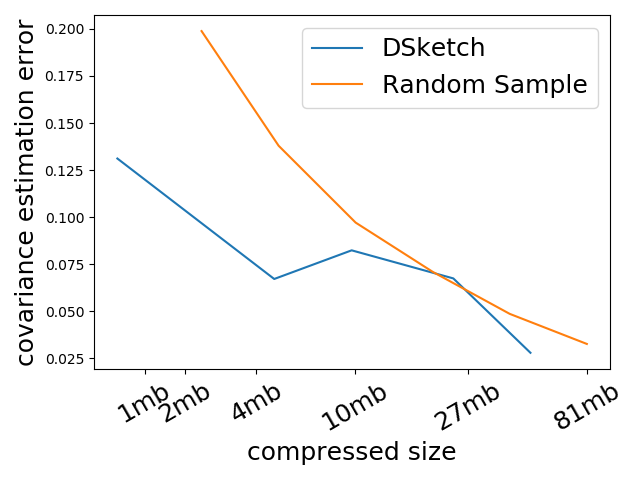}
  \caption{higgs }
\end{subfigure}
\begin{subfigure}{.24\textwidth}
  \centering
  \includegraphics[width=.98\linewidth]{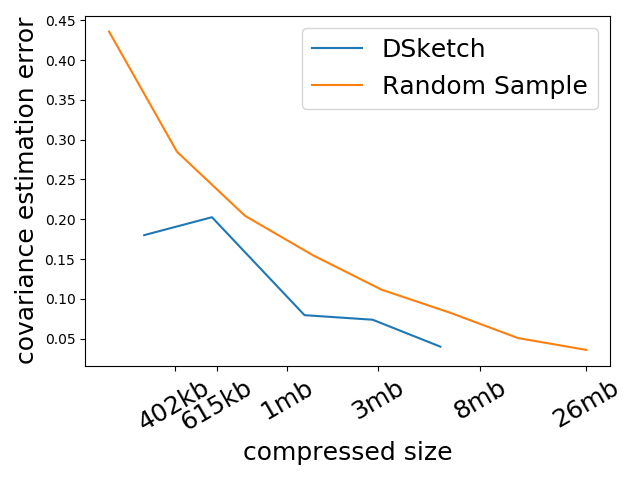}
  \caption{ susy}
\end{subfigure}
\begin{subfigure}{.24\textwidth}
  \centering
  \includegraphics[width=.98\linewidth]{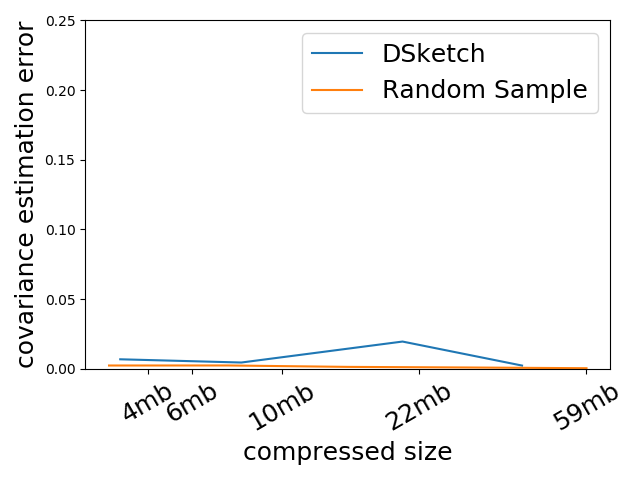}
  \caption{webspam(unigram)}
\end{subfigure}
\vspace{-0.2cm}
\caption{\textbf{Classification Results(top)} : Test accuracy of models trained on samples drawn from density sketches and a random sample. Models from Original data are horizontal lines. When compared for original accuracy, we get \textbf{$\approx$200x} Compression in "skin", \textbf{$\approx$150x} compression in "susy", \textbf{$\approx$50x} compression in "higgs" with Densiy Sketches. \textbf{Estimation Error(bottom)}: Mean Square Error of the predicted covariance matrix from samples drawn from Density Sketches and Random Sample as compared to the empirical covariance matrix of the original data. Datasets chosen as \ref{tab:datasets}.}
\label{fig:estimation}
\vspace{-0.4cm}
\end{figure*}

\vspace{-0.2cm}
\section{Experiments}
\vspace{-0.2cm}
\subsection{Visualization of samples from density sketches}
\vspace{-0.2cm}
Figure \ref{fig:visualization} (a) shows the samples drawn from the actual multi Gaussian distribution, whereas figure \ref{fig:visualization} (b) shows the samples drawn from the Density Sketch(DS) built on the samples from the true distribution. As can be seen, the two samples are indistinguishable. Figure \ref{fig:visualization} (c) shows some samples drawn from DS built over the "MNIST" dataset \cite{liblinear}. It can be seen that DS give sensible samples which can be identified as digits by the naked eye for different widths of partitions. In these experiments, we used l2-lsh random partitioning.

\vspace{-0.2cm}
\subsection{Evaluation of Samples on Classification Tasks}
\vspace{-0.2cm}

For most datasets, it is not possible to inspect samples visually. Hence we evaluate the quality of samples from DS by using them to train classification models.  

\textbf{Datasets:}
To perform a fair evaluation, we chose all datasets from the liblinear website \cite{liblinear}, which satisfy the constraints of 1) data dimension less than 500 and 2) the number of samples per class greater than 100,000. Large Datasets is the main application domain for DS. The datasets such obtained are noted in the table \ref{tab:datasets}. We use l2-lsh random partitions with 0.01 bandwidth

\textbf{Results:}
As can be seen, for "Higgs" dataset , the accuracy of the model achieved on original data of size 2.5GB, can be achieved by using a DS of size ~50MB. So we get around \textbf{50x compression} ! We see similar results for datasets of "skin" (\textbf{200x compression}, 670KB)  and "susy" (\textbf{150x compression}, 0.8GB) as well. The results show that DS is much more informative than Random Sample. The dimension of Webspam data is 254. The DS for this dataset is at a disadvantage due to Curse of Dimensionality : Number of datapoints required for good estimates is exponential in dimension. See Discussion \ref{sec:curse}  for qualitative discussion on this aspect. Due to these reasons the performance of DS is comparatively poor in "Webspam" dataset although it still beats Random Sampling in this experiment. 

\textbf{Other Baselines:} We consider comparing against other sophisticated baselines. However, none are comparable to DS in their purpose/utility. (1)Coresets for sampling would first require KDE estimation, which has a huge memory cost to construct the point set. Also, despite recent progress towards coresets in streaming setting \cite{phillips2020near}, coresets remain challenging to implement for  KDE problems~\cite{hbe}.(2) Algorithms like K-means clustering are inappropriate for streaming datasets and don't have mergeability property of DS. An alternative approach is to select points based on importance sampling~\cite{hbe}, geometric properties~\cite{cortes2016sparse}, and other sampling techniques~\cite{chen2012super}. However, recent experiments show that for many real-world datasets, random samples have a competitive performance with these approaches ~\cite{race}.  Dimensionality reduction via random projections is another way to reduce the size, however, given the relatively small dimension. It is unlikely to work well.

%\textit{\textbf{Coresets}}
%One might wonder why we have compared against random samples instead of sophisticated data summary techniques like coresets. In the past, it has been observed that a random sample is practically as good as coresets in higher dimensions \cite{hbe}. Hence, we use a random sample as a competing option. Also, we want to stress that generating samples from density sketches is a much more novel and richer approach to creating and using data summary than storing and using a random sample.
%\textit{\textbf{Feature hashing}}
%Feature hashing is yet another technique to reduce the memory foot print of the data. In the datasets we show, we cannot use feature hashing to get reasonable reduction in  size of the data due to relatively smaller dimension of data.
\vspace{-0.2cm}
\subsection{Estimation of statistical properties of dataset}
\vspace{-0.2cm}

We evaluate the DS on another task of evaluating different properties of the distribution on the same Datasets. Specifically, in this section, we show that samples drawn from the DS can be used effectively to estimate the Covariance Matrix of the underlying distribution. See Figure \ref{fig:estimation} for plots. The prediction from DS is superior to Random Sample except for "Webspam". Again, due to high dimensional data, DS do not perform that well. In this case it is worse than Random sample.

\vspace{-0.2cm}
\subsection{Manifestation of Theory in Experiments}
\vspace{-0.2cm}

As can be seen from theory, a lot of factors affect performance of DS such as $nh^d$, $ratio_h$, $(\#bins/KR)$ among others. In our experiments, we see that "Webspam" as higher $d$ and smaller $n$ as compared to, say, "higgs", hence the quality of histogram is bad for Webspam which translates to DS as well. Another factor that affects sampling distribution is cature ratio $ratio_h$, we observe that data is more scattered in "Webspam", hence is $ratio_h$ is worse than other datasets adversely affecting results.

\vspace{-0.2cm}
\section{Discussion: Curse of Dimensionality}\label{sec:curse}
\vspace{-0.2cm}

In the context of density estimators, the curse of dimensionality implies (1) that the number of data points required to get decent estimates of density increases exponentially with dimension.  (2) The number of bins in histograms is exponential. As Density Sketches are built over Histograms, they inherit this curse from Histograms.  With increased data collection, the issue of the unavailability of large amounts of data is fast vanishing.  We want to emphasize that Density Sketches' advantages are best seen when data is humongous. Density Sketches can absorb tons of data and give better density estimates and samples without increasing their memory usage. Also, most real data in high dimensions is clustered or stays on a low-dimensional manifold. Density Sketches, throw away empty bins, and only store the histogram's populated bins. Density Sketches can deal with the curse of dimensionality better than Histograms.

\vspace{-0.4cm}
\section{Conclusion}
\vspace{-0.2cm}
We introduce \textit{Density sketches (DS)\footnote{We provide code for Density Sketches in Supplementary}}: a sketch summarizing density from data samples. With very tiny size, streaming nature, sketch properties and capability to sample, DS has to potential to change the way we store, communicate and distribute data. 
\newpage
\bibliography{ref}
\bibliographystyle{unsrt}

\newpage
\onecolumn
\section{Appendix A}

\subsection{Theorem 2 : $F_H$ to $f$}
While estimating true distribution $f(x) : R^d \rightarrow R$, the integrated mean square error (IMSE) for the estimator $\hat{f_H}(x)$ using  regular histogram with width h and number of samples n, is 
\[
    IMSE(\hat{f_H}) \leq \frac{1}{nh^d} + \frac{R(f)}{n} + o(\frac{1}{n}) + \frac{h^2d}{4} R(\norm{\nabla f}{2})
\]
Specifically, its
\[
IV = \frac{1}{nh^d} + \frac{R(f)}{n} + o(\frac{1}{n})
\] and 
\[
ISB \leq \frac{h^2d}{4} R(\norm{\nabla f}{2})
\]
where $R(\phi)$ is the roughness of the function $\phi$ defined as $R(\phi) = \int \phi(x)^2 dx$

\begin{proof}

Let $x \in S$ .  S is the support of the distribution. The estimator $\hat{f_H}(x)$ is defined as, where V(x) is volume of bin in which x lies. Equivalently, we can also use V(b) to denote volume of bin b. For standard histogram, $V(x) = h^d$
\begin{equation}
    \hat{f_H}(x) = \frac{1}{n V(bin(x))} \Sigma_{i=1}^{n} \mathcal{I} (x_i \in bin(x))
\end{equation}

First let us consider the integrated variance.
\begin{equation}
    IV = \int_{x \in S} Var(\hat{f_H}(x)) dx  = \Sigma_{b \in bins(S) } \int_{x \in b} Var(\hat{f_H}(x)) dx 
\end{equation}
For a particular bin b, the variance is constant at all values of x. Also for a particular x in bin b, we can write the following for $Var(\hat{f_H}(x))$ using independence of samples.
\begin{equation}
    Var(\hat{f_H}(x)) = \frac{1}{nV(bin(x))^2} Var( \mathcal{I}(x_i \in bin(x))
\end{equation}
Also $Var( \mathcal{I}(x_i \in b)) = p_b ( 1-p_b)$ where $p_b$ is the probability of $x_i$ lying in bin b. That is, $p_b = \int_{x \in b} f(x) dx$

Using this in equation 2
\begin{equation}
    IV =  \Sigma_{b \in bins(S) } V(b) \frac{1}{nV(b)^2} p_b * (1-p_b)
\end{equation}
Simplifying, 
\begin{equation}
    IV =  \Sigma_{b \in bins(S)} \frac{1}{nV(b)} p_b * (1-p_b)
\end{equation}
For standard histogram V(b) is same across bins,
\begin{equation}
    IV =  \frac{1}{nV(b)} (\Sigma_{b \in bins(S)}  p_b  - \Sigma_{b \in bins(S)}  p_b^2) = \frac{1}{nV(b)} (1  - \Sigma_{b \in bins(S)}  p_b^2)
\end{equation}
Using mean value theorem, we can write, $p_b = V(b) f(\xi_b)$ for some point $\xi_b \in b$. 
\begin{equation}
\Sigma_{b \in bins} p_b^2 = \Sigma_{b \in bins} V(b)^2 f(\xi_b)^2 = V(b) \Sigma_{b \in bins} V(b) f(\xi_b)^2
\end{equation}

Using Rieman Integral approximation , we can write the following as the bin size reduces,
\begin{equation}
\Sigma_{b \in bins} V(b) f(\xi_b)^2 = \int_{x \in S} f^2(x) dx + o(1)
\end{equation}
$\int_{x \in S} f^2(x) dx $ is also known as the roughness of the function. Let us denote it using R(f). Hence
\begin{equation}
    IV = \frac{1}{nV(b)} (1  - V(b) (R(f) + o(1)))
\end{equation}
\begin{equation}
    IV = \frac{1}{nV(b)}   - \frac{R(f)}{n} + o(\frac{1}{n})))
\end{equation}
Putting $V(b) = h^d$
\begin{equation}
    IV = \frac{1}{nh^d}   - \frac{R(f)}{n} + o(\frac{1}{n})))
\end{equation}
Keeping only the leading term in the above expression,
\begin{equation}
    IV = O(\frac{1}{nh^d})
\end{equation}

Now let us look at the ISB for this estimator, $ISB(\hat{f_h}(x))$

\begin{equation}
    ISB(\hat{f_h}(x)) = \int_{x \in S} (E(\hat{f_H}(x) - f(x)))^2 dx
\end{equation}

Let us look at the estimator,
\begin{equation}
    \hat{f_H}(x) = \frac{1}{V(bin(x))} \int_{t \in bin(x)} f(t) dt
\end{equation}
Just to make it clear, $x \in R^d$, we will use it as a vector in the following. Using 2nd order multivariate taylor series expansion of this $f(t)$ around $x$, we get : 

\begin{equation}
    f(t) = f(x) + \ip{t-x}{\nabla f(x)} + \frac{1}{2} (t-x)^\top \mathcal{H}(f(x)) (t-x)
\end{equation}
Here $\mathcal{H}(f(t))$ is the hessian of f at t. Without loss of generality let us look at the $bin(x) = [0,h]^d$ that is the bin at the origin. Let us say it is $bin_0$
\begin{equation}
    \int_{t \in bin_0} f(t) dt = f(x) h^d + h^d \ip{ (\frac{h}{2} - x}{\nabla f(x)} + O(h^{d+2})
\end{equation}
where $x^{(j)}$ is the $j^{th}$ component of x.  Using eq 17 in eq 15, we get
\begin{equation}
    \hat{f_H}(x) = f(x)  + \ip{ (\frac{h}{2} - x)}{\nabla f(x)} + O(h^2)
\end{equation}
Hence, just keeping the leading term , we have
\begin{equation}
    Bias(\hat{f_H}(x)) =  \ip{ (\frac{h}{2} - x)}{\nabla f(x)}
\end{equation}
Now, 
\begin{equation}
    \int_{x \in b_0} Bias(\hat{f_H}(x))^2 dx =  \int_{x \in b_0} (\ip{ (\frac{h}{2} - x)}{\nabla f(x)})^2 dx
\end{equation}
Using cauchy's inequality, we get 
\begin{equation}
    \int_{x \in b_0} Bias(\hat{f_H}(x))^2 dx \leq  \int_{x \in b_0} \norm{(\frac{h}{2} - x)}{2}^2 \norm{\nabla f(x)}{2}^2 dx
\end{equation}
As $[h/2, h/2, ...h/2]$ is a mid point of the bin. The max norm of $x - h/2$ can be $h\sqrt{d}/2$
\begin{equation}
    \int_{x \in b_0} Bias(\hat{f_H}(x))^2 dx \leq \frac{h^2 d}{4} \int_{x \in b_0}\norm{\nabla f(x)}{2}^2 dx
\end{equation}

Now looking at ISB
\begin{equation}
    ISB = \Sigma_{b \in bins } \int_{x \in b_0} Bias(\hat{f_H}(x))^2 dx \leq  \frac{h^2 d}{4} \int_{x \in S} \norm{\nabla f(x)}{2}^2 dx
\end{equation}
\begin{equation}
    ISB \leq  \frac{h^2 d}{4} R(\norm{\nabla f}{2})
\end{equation}

\subsection{Theorem 3: $F_C$ to $f_H$}
While estimating true distribution $f(x) : R^d \rightarrow R$, the integrated mean square error (IMSE) for the estimator $\hat{f_C}(x)$ using  regular histogram with width h and number of samples n and countsketch with parameters (R:range, K:repetitions) and average-recovery, is 
\[
    IMSE(\hat{f_C}) = IMSE(\hat{f_H}) + \frac{\#bins}{KRnh^d}
\]
where $n_{nzp}$ is the number of non-zero partitions. Specifically, its 
\[
IV(\hat{f_C}) = IV(\hat{f_H}) + \frac{\#bins - 1}{KRnh^d} 
\] 
and 
\[
ISB(\hat{f_C}) = ISB(\hat{f_H})
\]
where $n_nzp$ is the number of non-zero bins/partitions.
\end{proof}

\begin{proof} 
Consider a Countsketch with range = R and just one repetition. Let it be parameterized by the randomly drawn hash functions $g: bin \longrightarrow \{0,1,2,...,R-1\}$ and $s: bin \longrightarrow \{-1,+1\}$. The estimate of density at point x can then be written as 

\begin{equation}
    \hat{f_C}(x) = \frac{1}{n V(bin(x))} ( c(bin(x)) + \Sigma_{i=1}^n  \mathcal{I}(x_i \notin bin(x)  \wedge g(bin(x_i)) == g(bin(x))) s(bin(x_i)) s(bin(x))
\end{equation}
We can rewrite this as ,
\begin{equation}
    \hat{f_C}(x) = \hat{f_H}(x)  + \frac{1}{n V(bin(x))} ( \Sigma_{i=1}^n  \mathcal{I}(x_i \notin bin(x) \wedge g(bin(x_i)) == g(bin(x))) s(bin(x_i)) s(bin(x))
\end{equation}

where c(.) is count and V(.) is volume of the bins. As E(s(b)) = 0, it can be clearly seen that. 
\begin{equation}
E(\hat{f_C}(x)) =  E(\hat{f_H}(x))
\end{equation}
Hence, it follows that 
\begin{equation}
    ISB(\hat{f_C}(x))  = ISB(\hat{f_H}(x))
\end{equation}
It can be checked that each of the terms in the summation for right hand side of equation 26 including the terms in $\hat{f_H}(x)$ are independent to each other . i.e. covariance between them is 0. Hence we can write the variance of our estimator as,
\begin{equation}
    Var(\hat{f_C}(x)) = Var(\hat{f_H}(x)) + \frac{1}{n V^2(bin(x))} Var (\mathcal{I}(x_i \notin bin(x) \wedge g(bin(x_i)) == g(bin(x))) s(bin(x_i)) s(bin(x)))
\end{equation}
\begin{equation}
    Var(\hat{f_C}(x)) = Var(\hat{f_H}(x)) + \frac{1}{n V^2(bin(x))} E (\mathcal{I}(x_i \notin bin(x) \wedge g(bin(x_i)) == g(bin(x))))^2
\end{equation}
\begin{equation}
    Var(\hat{f_C}(x)) = Var(\hat{f_H}(x)) + \frac{1}{n V^2(bin(x))} (1 - p_{bin(x)}) \frac{1}{R})
\end{equation}

Hence, IV is
\begin{equation}
    IV(\hat{f_C}(x)) = IV(\hat{f_H}(x)) + \int_{x \in S}  \frac{1}{n V^2(bin(x))} (1 - p_{bin(x)}) \frac{1}{R})
\end{equation}
\begin{equation}
    IV(\hat{f_C}(x)) = IV(\hat{f_H}(x)) + \Sigma_{b \in bins} \int_{x \in b}  \frac{1}{n V^2(b)} (1 - p_{b}) \frac{1}{R})
\end{equation}
\begin{equation}
    IV(\hat{f_C}(x)) = IV(\hat{f_H}(x)) + \Sigma_{b \in bins}  \frac{1}{n V(b)} (1 - p_{b}) \frac{1}{R})
\end{equation}
Assuming standard partitions. $V(b) = h^d$ for all b
\begin{equation}
    IV(\hat{f_C}(x)) = IV(\hat{f_H}(x)) +  \frac{1}{n h^d}  \frac{(\#bins - 1)}{R}
\end{equation}
With average recovery, with K repetitions, the analysis can be easily extended to get IV as
\begin{equation}
    IV(\hat{f_C}(x)) = IV(\hat{f_h}(x)) +  \frac{1}{n h^d}  \frac{(\#bins - 1)}{KR}
\end{equation}
The ISB remains same in this case.
\end{proof}

\subsection{Theorem 4: $f^*_C to f_C$ }
While estimating true distribution $f(x) : R^d \rightarrow R$, the integrated mean square error (IMSE) for the estimator $\hat{f^*_C}(x)$ using  regular histogram with width h and number of samples n and countsketch with parameters (R:range, K:repetitions), is related to the estimator $\hat{f_C}(x)$ as follows 
\[
    IMSE(\hat{f_C(x)}) - \epsilon (N + 2M) \leq IMSE(\hat{f^*_C(x)}) \leq IMSE(\hat{f_C(x)}) + \epsilon (N + 2M)
\]
Specifically, 
\[
IV(\hat{f_C(x)}) - 2 \epsilon M \leq IV(\hat{f^*_C(x)}) \leq IV(\hat{f_C(x)}) + 2 \epsilon M
\] 
and 
\[
ISB(\hat{f_C(x)}) - \epsilon N \leq ISB(\hat{f^*_C(x)}) \leq ISB(\hat{f_C(x)}) + \epsilon N
\]
where 
\[
M \leq  IV(\hat{f_C}(x)) + 2 (R(f) + \frac{h^2d}{4} R(\norm{\nabla f}{2}) + h\sqrt{d} \int_{x \in S}(f(x) \norm{\nabla f}{2}) )
\]
\[
N = (1 + ISB(\hat{f_C}(x)))
\]
with probability $(1 - \delta)$ where $\delta = \frac{\#bins }{\epsilon^2 n R}$

\begin{proof}
Let us look at the estimator 
\begin{equation}
    \hat{f^*_C(x)} = \frac{\widehat{c(bin(x))}}{V(bin(x)) \Sigma_b \widehat{c(b)}} = \hat{f_C(x)} * \frac{n}{\hat{n}}
\end{equation}
where $\hat{n} = \Sigma_b \widehat{c(b)}$ and $n = \Sigma_b c(b)$
\end{proof}
\paragraph{$\hat{\mathbf{n}}$ and its relation to n}
Let us first analyse $\hat{n}$ and how it is related to n. 
\begin{equation}
    \hat{n} = \Sigma_b \widehat{c(b)} = \Sigma_b \Sigma_{i=1}^n \mathcal{I}(x_i \in b) + \mathcal{I}(x_i \notin b \wedge g(bin(x_i)) == g(b)) s(bin(x_i)) s(b)
\end{equation}

\begin{equation}
    \hat{n} = \Sigma_{b,i}  \mathcal{I}(x_i \in b) + \mathcal{I}(x_i \notin b \wedge g(bin(x_i)) == g(b)) s(bin(x_i)) s(b)
\end{equation}
Note that $E(\hat{n}) = n$. For varaince, observe that most of the terms in the summation have covariance 0, except the terms $Cov(\mathcal{I}(x_i \in b_1), \mathcal{I}(x_i \in b_2))$ which are negatively correlated. Hence
\begin{equation}
    \begin{split}
Var(\hat{n}) = &\Sigma_{b,i} Var( \mathcal{I}(x_i \in b)) + Var(\mathcal{I}(x_i \notin b \wedge g(bin(x_i)) != g(b)) s(bin(x_i)) s(b)) + \\ 
& 2 \Sigma_{i, b_1, b_2, b_1 \neq b_2} Cov(\mathcal{I}(x_i \in b_1), \mathcal{I}(x_i \in b_2))        
    \end{split}
\end{equation}
We know that 
\begin{align*}
    & Var( \mathcal{I}(x_i \in b)) = p_b ( 1 - p_b) \\
    & Var(\mathcal{I}(x_i \notin b \wedge g(bin(x_i)) == g(b)) s(bin(x_i)) s(b)) = E(\mathcal{I}(x_i \notin b \wedge g(bin(x_i)) != g(b))^2) = \frac{1-p_b}{R}\\
    & Cov(\mathcal{I}(x_i \in b_1), \mathcal{I}(x_i \in b_2)) =  - p_{b_1} p_{b_2}
\end{align*}
Hence, we pluggin in the values in previous equation , 
\begin{equation}
Var(\hat{n}) = n \Sigma_b  p_b(1-p_b) + n \Sigma_{b} \frac{1-p_b}{R} - 2 n \Sigma_{b_1, b_2, b_1\neq b_2} p_{b_1} p_{b_2}
\end{equation}
\begin{equation}
Var(\hat{n}) = n (1 - \Sigma_b p_b^2)   + n \Sigma_{b} \frac{1-p_b}{R} - 2 n \Sigma_{b_1, b_2} p_{b_1} p_{b_2}
\end{equation}
\begin{equation}
Var(\hat{n}) = n \{ (1 + \Sigma_{b} \frac{1-p_b}{R} - (\Sigma_b p_b^2)   - 2 n \Sigma_{b_1, b_2} p_{b_1} p_{b_2} )\}
\end{equation}
\begin{equation}
Var(\hat{n}) = n \{ (1 + \Sigma_{b} \frac{1-p_b}{R} - (\Sigma_b p_b)^2\}
\end{equation}
\begin{equation}
Var(\hat{n}) = n \{ \Sigma_{b} \frac{1-p_b}{R} \}
\end{equation}
\begin{equation}
Var(\hat{n}) = \frac{n (\#bins - 1)}{R} < \frac{n (\#bins)}{R}
\end{equation}
Using Chebyshev's inequality , we have 
\begin{equation}
P ( | \hat{n} - n | > \epsilon n ) \leq \frac{Var(\hat{n})}{\epsilon^2 n^2}
\end{equation}

\begin{equation}
P ( | \hat{n} - n | > \epsilon n ) \leq \frac{\#bins }{\epsilon^2 n R}
\end{equation}
Hence with probability $(1 - \delta)$, $\delta = \frac{\#bins }{\epsilon^2 n R}$, $\hat{n}$ is within $\epsilon$ multiplicative error. 

\paragraph{relation of pointwise Bias and ISB}
With probability $1-\delta$,
\begin{equation}
    \frac{\hat{f_C}(x)}{1+\epsilon} \leq \hat{f^*_C(x)} \leq \frac{\hat{f_C(x)}}{1-\epsilon}
\end{equation}
As expectations respect inequalities
\begin{equation}
    \frac{E(\hat{f_C}(x))}{1+\epsilon} \leq E(\hat{f^*_C(x)}) \leq \frac{E(\hat{f_C(x)})}{1-\epsilon}
\end{equation}
\begin{equation}
    \frac{E(\hat{f_C}(x))}{1+\epsilon} - f(x) \leq Bias(\hat{f^*_C(x)}) \leq \frac{E(\hat{f_C(x)})}{1-\epsilon} -f(x)
\end{equation}
\begin{equation}
    \frac{Bias(\hat{f_C}(x)) - \epsilon f(x) }{1+\epsilon} \leq Bias(\hat{f^*_C(x)}) \leq \frac{Bias(\hat{f_C(x)}) + \epsilon f(x)}{1-\epsilon}
\end{equation}
\begin{equation}
    \frac{Bias(\hat{f_C}(x)) - \epsilon f(x) }{1+\epsilon} \leq Bias(\hat{f^*_C(x)}) \leq \frac{Bias(\hat{f_C(x)}) + \epsilon f(x)}{1-\epsilon}
\end{equation}
Integrating expressions again respects inequalities

\begin{equation}
    \frac {ISB(\hat{f_C}(x)) - \epsilon \int f(x) }{1+\epsilon} \leq ISB(\hat{f^*_C(x)}) \leq \frac{ISB(\hat{f_C(x)}) + \epsilon \int f(x)}{1-\epsilon}
\end{equation}

\begin{equation}
    \frac {ISB(\hat{f_C}(x)) - \epsilon }{1+\epsilon} \leq ISB(\hat{f^*_C(x)}) \leq \frac{ISB(\hat{f_C(x)}) + \epsilon}{1-\epsilon}
\end{equation}
Using first order taylor expansion of $\frac{1}{1 + \epsilon}$ and ignore square terms
\begin{equation}
    (1-\epsilon) ISB(\hat{f_C}(x)) - \epsilon  \leq ISB(\hat{f^*_C(x)}) \leq (1+\epsilon) ISB(\hat{f_C(x)}) + \epsilon
\end{equation}
\begin{equation}
    ISB(\hat{f_C}(x)) - \epsilon (1 + ISB(\hat{f_C}(x)))  \leq ISB(\hat{f^*_C(x)}) \leq ISB(\hat{f_C(x)}) + \epsilon (1 + ISB(\hat{f_C}(x)))
\end{equation}
Hence, 
\begin{equation}
    ISB(\hat{f_C}(x)) - \epsilon N   \leq ISB(\hat{f^*_C(x)}) \leq ISB(\hat{f_C(x)}) + \epsilon N
\end{equation}
where
\begin{equation*}
    N = (1 + ISB(\hat{f_C}(x)))
\end{equation*}

\paragraph{Point wise variance and IV}
Using the similar arguments 

\begin{equation}
    \frac{E(\hat{f_C}^2(x))}{(1+\epsilon)^2} - \frac{E^2(\hat{f_C}(x))}{(1-\epsilon)^2} \leq Var(\hat{f^*_C}(x)) \leq \frac{E(\hat{f_C}^2(x))}{(1-\epsilon)^2} - \frac{E^2(\hat{f_C}(x))}{(1+\epsilon)^2}
\end{equation}

Again making first order taylor expansions of denominator and ignoring square terms

\begin{equation}
    Var(\hat{f_C}(x))  -2 \epsilon (E(\hat{f_C}^2(x) + E^2(\hat{f_C}(x))) \leq Var(\hat{f^*_C}(x)) \leq Var(\hat{f_C}(x)) + 2(E(\hat{f_C}^2(x) + E^2(\hat{f_C}(x)))
\end{equation}

Since, $Var(\hat{f_C(x)}) = E(\hat{f_C}^2(x)) - E^2(\hat{f_C}(x))$
\begin{equation}
    Var(\hat{f_C}(x))  -2 \epsilon ( Var(\hat{f_C}(x))  + 2 E^2(\hat{f_C}(x))) \leq Var(\hat{f^*_C}(x)) \leq Var(\hat{f_C}(x)) + 2 \epsilon( Var(\hat{f_C}(x)) + 2 E^2(\hat{f_C}(x)))
\end{equation}
\begin{equation}
    IV(\hat{f_C}(x))  -2 \epsilon ( IV(\hat{f_C}(x))  + 2 \int_{x \in S} E^2(\hat{f_C}(x))) \leq IV(\hat{f^*_C}(x)) \leq IV(\hat{f_C}(x)) + 2 \epsilon( IV(\hat{f_C}(x)) + 2 \int_{x \in S} E^2(\hat{f_C}(x)))
\end{equation}
Let us now figure out the  $\int_{x \in S} E^2(\hat{f_C}(x))$

\begin{equation}
\int_{x \in S} E^2(\hat{f_C}(x)) = \int_{x \in S} E^2(\hat{f_H}(x))
\end{equation}
From equation 18,  E($\hat{f_H}(x))^2 = f(x)^2  + (\ip{ (\frac{h}{2} - x)}{\nabla f(x)} )^2 + 2f(x) \ip{ (\frac{h}{2} - x)}{\nabla f(x)}  $
\begin{equation}
\int_{x \in S} E^2(\hat{f_H}(x)) \leq R(f) + \frac{h^2d}{4} R(\norm{\nabla f}{2}) + h\sqrt{d} \int_{x \in S}(f(x) \norm{\nabla f}{2}) 
\end{equation}

Hence, 
\begin{equation}
    IV(\hat{f_C}(x))  -2 \epsilon M \leq IV(\hat{f^*_C}(x)) \leq IV(\hat{f_C}(x)) + 2 \epsilon M 
\end{equation}

Where
\begin{equation}
M \leq  IV(\hat{f_C}(x)) + 2 (R(f) + \frac{h^2d}{4} R(\norm{\nabla f}{2}) + h\sqrt{d} \int_{x \in S}(f(x) \norm{\nabla f}{2}) )
\end{equation}

\subsection{Lemma 1}
Estimators $\hat{f_S}(x)$ and $\hat{f^*_C}(x)$,  obtained from the Density Sketch with parameters(R,K,H) using histogram of width h built over n i.i.d samples drawn from true distribution have a relation
\[
\int |\hat{f^*_C}(x) - \hat{f_S}(x)| dx = 2(1-ratio_h)
\]
where $ratio_h$ is the capture ratio as defined in section 3
\begin{equation}
\int |\hat{f^*_C}(x) - \hat{f_S}(x)| dx = \Sigma_{b \in bins}     \int_{x \in b} |\hat{f^*_C}(x) - \hat{f_S}(x)| dx
\end{equation}

\begin{equation}
\int |\hat{f^*_C}(x) - \hat{f_S}(x)| dx = \Sigma_{b \in bins(H)}     \int_{x \in b} |\hat{f^*_C}(x) - \hat{f_S}(x)| dx + \Sigma_{b \notin bins(H)}     \int_{x \in b} |\hat{f^*_C}(x) - \hat{f_S}(x)| dx
\end{equation}

we know that for $x \in b , b \notin bins(H)$, $\hat{f_S}(x)$ = 0. Hence, 

\begin{equation}
\int |\hat{f^*_C}(x) - \hat{f_S}(x)| dx = \Sigma_{b \in bins(H)}     \int_{x \in b} |\hat{f^*_C}(x) - \hat{f_S}(x)| dx + \Sigma_{b \notin bins(H)}     \int_{x \in b} \hat{f^*_C}(x) dx
\end{equation}
$\int_{x \in b} \hat{f^*_C}(x) dx$ is the probability of a data point lying in that bucket according to $\hat{f^*_C}(x)$

\begin{equation}
\int |\hat{f^*_C}(x) - \hat{f_S}(x)| dx = \Sigma_{b \in bins(H)}     \int_{x \in b} |\hat{f^*_C}(x) - \hat{f_S}(x)| dx + \Sigma_{b \notin bins(H)}  \frac{\hat{c_b}}{\hat{n}}
\end{equation}
For points $x \in b , b \in bins(H)$, $\hat{f^*_C}(x) * \hat{n} = \hat{f_S}(x) * \hat{n_h}$, Hence, $\hat{f_S}(x) = \frac{\hat{n}}{\hat{n_h}} \hat{f^*_C}(x) $
\begin{equation}
\int |\hat{f^*_C}(x) - \hat{f_S}(x)| dx = \Sigma_{b \in bins(H)}     \int_{x \in b} \hat{f^*_C}(x) (\frac{\hat{n}}{\hat{n_h}} - 1) dx + \Sigma_{b \notin bins(H)}  \frac{\hat{c_b}}{\hat{n}}
\end{equation}
\begin{equation}
\int |\hat{f^*_C}(x) - \hat{f_S}(x)| dx = \Sigma_{b \in bins(H)}     \int_{x \in b} \hat{f^*_C}(x) (\frac{\hat{n}}{\hat{n_h}} - 1) dx + \Sigma_{b \notin bins(H)}  \frac{\hat{c_b}}{\hat{n}}
\end{equation}

\begin{equation}
\int |\hat{f^*_C}(x) - \hat{f_S}(x)| dx = (\frac{\hat{n}}{\hat{n_h}} - 1) \Sigma_{b \in bins(H)}  \frac{\hat{c_b}}{\hat{n}} + \Sigma_{b \notin bins(H)}  \frac{\hat{c_b}}{\hat{n}}
\end{equation}
\begin{equation}
\int |\hat{f^*_C}(x) - \hat{f_S}(x)| dx = (\frac{\hat{n}}{\hat{n_h}} - 1) (\frac{\hat{n_h}}{\hat{n}}) +  \frac{\hat{n} - \hat{n_h}}{\hat{n}}
\end{equation}

\begin{equation}
\int |\hat{f^*_C}(x) - \hat{f_S}(x)| dx = (1 - \frac{\hat{n_h}}{\hat{n}}) +  \frac{\hat{n} - \hat{n_h}}{\hat{n}}
\end{equation}

\begin{equation}
\int |\hat{f^*_C}(x) - \hat{f_S}(x)| dx = 2 (1 - \frac{\hat{n_h}}{\hat{n}})
\end{equation}
\begin{equation}
\int |\hat{f^*_C}(x) - \hat{f_S}(x)| dx = 2 (1 - ratio_h)
\end{equation}

\subsection{Theorem 5}
The IMSE of estimator $\hat{f_S}(x)$ obtained from the Density Sketch with parameters(R,K,H) using histogram of width h built over n i.i.d samples drawn from true distribution f(x) is 
\[
IMSE(\hat{f_S}(x)) \leq 12(1-ratio_h)^2 + 3 IMSE(\hat{f^*_C}(x))
\]
where $ratio_h$ is the capture ratio as defined in
\begin{proof}
Giving a very loose relation between $\hat{f_S}$ and f. We can write
\begin{align}
    &\int (\hat{f_S}(x)  - f(x))^2 dx = \int ((\hat{f_S}(x) -\hat{f^*_C}(x))  -  (\hat{f^*_C}(x)- f(x)))^2 dx \\
    &\int (\hat{f_S}(x)  - f(x))^2 dx \leq 3 \int (\hat{f_S}(x) -\hat{f^*_C}(x))^2 dx +  3 \int (\hat{f^*_C}(x) -f(x))^2 dx\\
    &\int (\hat{f_S}(x)  - f(x))^2 dx \leq 3 (\int |(\hat{f_S}(x) -\hat{f^*_C}(x))|dx)^2 +  3 \int (\hat{f^*_C}(x) -f(x))^2 dx\\
    &\int (\hat{f_S}(x)  - f(x))^2 dx \leq 12 (1 - ratio_h)^2 +  3 \int (\hat{f^*_C}(x) -f(x))^2 dx
\end{align}
\begin{equation}
    IMSE = MISE(\hat{f_S}(x)) \leq 12(1-ratio_h)^2 + 3 IMSE(\hat{f^*_C}(x))
\end{equation}
\end{proof}

\section{Theorem 1 (Main Theorem) combines all other theorems}
This theorem directly relates the distribution $\hat{f_S}(x)$ to the true distribution. f(x)
\begin{equation}
IMSE(\hat{f_S}(x)) \leq 12(1-ratio_h)^2 + 3 IMSE(\hat{f^*_C}(x))
\end{equation}

\begin{equation}
IMSE(\hat{f_S}(x)) \leq 12(1-ratio_h)^2 + 3 (IMSE(\hat{f_C}(x) + \epsilon (N+2M)))
\end{equation}

\begin{equation}
IMSE(\hat{f_S}(x)) \leq 12(1-ratio_h)^2 + 3 (IMSE(\hat{f_H}) + \frac{\#bins - 1}{KRnh^d} + \epsilon (N+2M)))
\end{equation}

\begin{equation}
IMSE(\hat{f_S}(x)) \leq 12(1-ratio_h)^2 + 3 (\frac{1}{nh^d} + \frac{R(f)}{n} + o(\frac{1}{n}) + \frac{h^2d}{4} R(\norm{\nabla f}{2}) + \frac{\#bins - 1}{KRnh^d} + \epsilon (N+2M)))
\end{equation}
     
\begin{align*}
N = (1 + ISB(\hat{f_C}))\\    
N \leq 1 +  \frac{h^2d}{4} \mathcal{R}(\norm{\nabla f}{2})
\end{align*}

\begin{align*}
M \leq  IV(\hat{f_C}) + 2 \mathcal{R}(f) + \frac{h^2d}{4} \mathcal{R}(\norm{\nabla f}{2}) + h\sqrt{d} \int_{x \in S}(f(x) \norm{\nabla f}{2}) )    \\
M \leq  IV(\hat{f_H}) + \frac{\#bins - 1}{KRnh^d}  + 2 \mathcal{R}(f) + \frac{h^2d}{4} \mathcal{R}(\norm{\nabla f}{2}) + h\sqrt{d} \int_{x \in S}(f(x) \norm{\nabla f}{2}) )    \\
M \leq  \frac{1}{nh^d} + \frac{\mathcal{R}(f)}{n} + o(\frac{1}{n}) + \frac{\#bins - 1}{KRnh^d}  + 2 \mathcal{R}(f) + \frac{h^2d}{4} \mathcal{R}(\norm{\nabla f}{2}) + h\sqrt{d} \int_{x \in S}(f(x) \norm{\nabla f}{2}) )    
\end{align*}

\begin{equation}
IMSE(\hat{f_S}(x)) \leq 12(1-ratio_h)^2 + 3 (\frac{1}{nh^d} + \frac{R(f)}{n} + o(\frac{1}{n}) + (1+ \epsilon) \frac{h^2d}{4} R(\norm{\nabla f}{2}) + \frac{\#bins - 1}{KRnh^d} + 2 \epsilon M + \epsilon)
\end{equation}

\begin{align*}
IMSE(\hat{f_S}(x)) \leq &12(1-ratio_h)^2 +\\ &3 (1+2\epsilon) (\frac{1}{nh^d} + \frac{R(f)}{n} + o(\frac{1}{n}) + \frac{\#bins - 1}{KRnh^d}) + \\
&3(1+ 3\epsilon) \frac{h^2d}{4} R(\norm{\nabla f}{2}) +\\
 &3 \epsilon (1 + 2 \mathcal{R}(f) + h\sqrt{d} \int_{x \in S}(f(x) \norm{\nabla f}{2})))
\end{align*}.

\section{Other Base lines}

\textit{Coresets:} We considered a comparison with sophisticated data summaries such as coresets. Briefly, a coreset is a collection of (possibly weighted) points that can be used to estimate functions over the dataset. To use coresets to generate a synthetic dataset, we would need to estimate the KDE. Unfortunately, coresets for the KDE suffer from practical issues such as a large memory cost to construct the point set. Despite recent progress toward coresets in the streaming environment~\cite{phillips2020near}, coresets remain difficult to implement for real-world KDE problems~\cite{hbe}. 

\textit{Clustering and Importance Sampling:} Another reasonable strategy is to represent the dataset as a collection of weighted cluster centers, which may be used to compute the KDE and sample synthetic points. Unfortunately, algorithms such as $k$-means clustering are inappropriate for large streaming datasets and do not have the same mergeability properties as our sketch. Furthermore, such techniques are unlikely to substantially improve over random sampling when the samples is spread sufficiently well over the support of the distribution. An alternative approach is to select points from the dataset based on importance sampling~\cite{hbe}, geometric properties~\cite{cortes2016sparse}, and 
other sampling techniques~\cite{chen2012super}. However, recent experiments show that for many real-world datasets, random samples have competitive performance when compared to point sets obtained via importance sampling and cluster-based approaches~\cite{race}.

\textit{Dimensionality Reduction:} One can also apply sketching algorithms to compress a dataset by reducing the dimension of each data point via feature hashing, random projections or similar methods~\cite{achlioptas2003database}. 
However, this is unlikely to perform well in our evaluation since our datasets are already relatively low-dimensional. Such algorithms also fail to address the streaming setting, where $N$ can grow very large, because the size of the compressed representation is linear in $N$. Finally, most dimensionality reduction algorithms do not easily permit the generation of more synthetic data in the original metric space.

\end{document}